\documentclass[aps,twocolumn,preprintnumbers,amsmath,amssymb,nofootinbib,superscriptaddress,notitlepage]{revtex4-1}

\usepackage{slashed}
\usepackage{epsfig}
\usepackage[utf8]{inputenc}
\usepackage{txfonts}

\usepackage{color}

\begin{document}

\title{Axial meson exchange and the $Z_c(3900)$ and $Z_{cs}(3985)$
  resonances \\ as heavy hadron molecules}

\author{Mao-Jun Yan}
\affiliation{School of Physics,  Beihang University, Beijing 100191, China}

\author{Fang-Zheng Peng}
\affiliation{School of Physics,  Beihang University, Beijing 100191, China}

\author{Mario {S\'anchez S\'anchez}}
\affiliation{Centre d'\'Etudes Nucl\'eaires, CNRS/IN2P3, Universit\'e de Bordeaux, 33175 Gradignan, France}

\author{Manuel Pavon Valderrama}\email{mpavon@buaa.edu.cn}
\affiliation{School of Physics, Beihang University, Beijing 100191, China}

\date{\today}


\begin{abstract} 
  \rule{0ex}{3ex}
  Early speculations about the existence of heavy hadron molecules
  were grounded on the idea that light-meson exchanges forces
  could lead to binding.
  In analogy to the deuteron, the light-mesons usually considered include
  the pion, sigma, rho and omega, but not the axial meson $a_1(1260)$.
  Though it has been argued in the past that the coupling of the axial meson
  to the nucleons is indeed strong, its mass is considerably heavier than
  that of the vector mesons and thus its exchange ends up being suppressed.
  Yet, this is not necessarily the case in heavy hadrons molecules:
  we find that even though the contribution to binding from the axial meson
  is modest, it cannot be neglected in the isovector sector
  where vector meson exchange cancels out.
  This might provide a natural binding mechanism for molecular candidates such
  as the $Z_c(3900)$, $Z_c(4020)$ or the more recently observed $Z_{cs}(3985)$.
  However the $Z_{cs}(3985)$ is more dependent on a mixture of different
  factors, which (besides axial meson exchange) include $\eta$ exchange
  and the nature of scalar meson exchange. Together they point
  towards the existence of two $Z_{cs}(3985)$-like resonances
  instead of one, while the observations about the role of scalar
  meson exchange in the $Z_{cs}(3985)$ might be relevant
  for the $P_{cs}(4459)$.
  Finally, the combination of axial meson exchange and flavor symmetry
  breaking effects indicates that the isovector $J^{PC} = 0^{++}$ $D^*\bar{D}^*$
  and the strange $J^P = 2^{+}$ $D^*\bar{D}_s^*$ molecules are
  the most attractive configurations and thus the most
  likely molecular partners of the $Z_c(3900)$, $Z_c(4020)$
  and $Z_{cs}(3985)$.
\end{abstract}

\maketitle

\section{Introduction}

Heavy hadron molecules were originally theorized as
an analogy to the deuteron~\cite{Voloshin:1976ap,DeRujula:1976qd}.
The argument is that the same type of forces binding two nucleons together
might bind other hadrons as well.
Since then a continuous inflow of ideas from nuclear physics has enriched
our understanding of heavy molecular states, ranging from phenomenological
approaches such as light-meson exchanges~\cite{Tornqvist:1991ks,Tornqvist:1993ng,Ericson:1993wy,Liu:2008fh} 
to modern effective field theory (EFT) formulations~\cite{Braaten:2003he,Fleming:2007rp,Mehen:2011yh,Valderrama:2012jv,Nieves:2012tt}.
This is not at all surprising: in both cases we are dealing with hadrons,
where nucleons happen to be the most well-studied of all hadrons.

Yet the origin and derivation of nuclear forces has itself a tortuous
and winding history, in which many competing ideas have been proposed
but few have succeeded~\cite{Machleidt:2001rw}.
The reasons behind the failures are important though, as they might
be specific to nucleons.
If we focus on light-meson exchange forces, the idea is that the nuclear
potential can be derived from the exchange of a few light mesons,
which usually include the pion, the sigma, the rho and the omega,
i.e. the one boson exchange model
(OBE)~\cite{Machleidt:1987hj,Machleidt:1989tm}.
Mesons heavier than the nucleon are generally not expected to have a sizable
contribution to the nuclear force: their Compton wavelength is shorter than
the size of the nucleon and the forces generated by their exchange
are heavily suppressed.

A prominent example is the axial meson $a_1(1260)$, which is expected
to have a considerably strong coupling to
the nucleons~\cite{Durso:1984um,Stoks:1996yj,Coon:1996wj}.
It is also heavier than the vector mesons: the ratio of the masses of
the axial and rho mesons, $m_{a1}$ and $m_{\rho}$, is $m_{a_1} / m_{\rho} \sim 1.6$.
In fact it is even heavier than the nucleon and its influence
on the description of the nuclear force has turned out to be
rather limited~\cite{Cordon:2010ez}.
However this is not necessarily the case for heavy hadrons: on the one hand,
they are heavier than axial mesons and nucleons, and on the other vector
meson exchange cancels out in a few specific molecular configurations,
which increases the relative importance of the axial meson.

The axial meson has a particularly interesting feature: its quantum numbers
$I^G (J^{PC}) = 1^{-} (1^{++})$ 
indicate that it can mix with the axial current of the pions.
That is, we can modify the axial pion current by including a term proportional
to the axial meson
\begin{eqnarray}
  \partial_{\mu} \pi \to
  \partial_{\mu} \pi + \lambda_1 \, m_{a1} \, a_{1 \mu} \, ,
  \label{eq:axial-current}
\end{eqnarray}
where $\pi$ and $a_{1\mu}$ are the pion and axial meson fields
and $\lambda_1$ a proportionality constant, which we expect
to be in the $\lambda_1 \sim (1.6-2.1)$ range.
From this the coupling of the axial meson with the charmed hadrons $D$ and
$D^*$ will be proportional to their axial coupling to
the pions, $g_1$:
\begin{eqnarray}
  g_{\,\rm Y(a_1)} = \frac{1}{\sqrt{3}}\,\lambda_1\,\frac{m_{a1}}{f_{\pi}}\,g_1
  \sim (8-11)\, g_1 \, ,
  \label{eq:relative-strength}
\end{eqnarray}
where the coupling is defined by matching the axial meson exchange
potential to a Yukawa (check Eqs.~(\ref{eq:V-a1}) and (\ref{eq:Yukawa-like})).
Depending on the configuration of the two-hadron system under consideration,
this exchange potential might be remarkably attractive and even explain
binding if other light-meson exchanges are suppressed.

The reason why we are interested in the axial meson is a specific difficulty
when explaining the isovector hidden-charm $Z_c(3900)$~\cite{Ablikim:2013mio},
$Z_c^*(4020)$~\cite{Ablikim:2013wzq} and $Z_{cs}(3985)$~\cite{Ablikim:2020hsk}
resonances as hadronic molecules:
while their closeness to the $D^*\bar{D}$, $D^*\bar{D}^*$ and
$D^*\bar{D}_s$-$D\bar{D}_s^*$ thresholds suggest the molecular nature
of the $Z_c$'s~\cite{Wang:2013cya,Guo:2013sya,Voloshin:2013dpa,Liu:2013vfa,Albaladejo:2015lob} and the $Z_{cs}$~\cite{Yang:2020nrt,Cao:2020cfx,Du:2020vwb,Sun:2020hjw,Wang:2020htx,Ikeno:2020csu}, the reasons why their interaction
is strong remains elusive.
It happens that rho- and omega-exchange cancel out for the $Z_c$ and $Z_c^*$,
which in turn requires a binding mechanism not involving the vector mesons
(where this cancellation does not happen with axial mesons, as we will see).
Possible explanations include one-pion and sigma
exchange~\cite{Liu:2008tn,Sun:2011uh,Liu:2019stu},
two-pion exchanges (the correlated part of which
is sometimes interpreted as sigma exchange) and charmonium
exchanges~\cite{Aceti:2014kja,Aceti:2014uea,He:2015mja,Dong:2021juy}.
Here we will investigate how axial meson exchange works
as a binding mechanism within the OBE model.

The manuscript is structured as follows: in Sect.~\ref{sec:axial} we will
derive the axial meson exchange potential for the charmed meson-antimeson
system. In Sect.~\ref{sec:vector} we will review scalar and vector meson
exchange and the potentials they generate, which are still an important
part of the OBE potential. In Sect.~\ref{sec:Zc} we will investigate how
the inclusion of the axial meson makes the simultaneous description of
the $X(3872)$ and the $Z_c(3900)$ more compatible with each other.
In Sect.~\ref{sec:Zcs} we will consider how the previous ideas apply
to the $Z_{cs}(3985)$ and how the OBE potential can be made compatible
with the expectations from SU(3)-flavor symmetry.
Finally in Sect.~\ref{sec:conclusions} we will explain our conclusions.

\section{Axial meson exchange}
\label{sec:axial}

First we will derive the potential generated by axial meson exchange.
We will begin with the interaction Lagrangian between heavy mesons and pions
and from it we will derive the Lagrangian and potential for axial mesons.

The quark content of the heavy mesons is $Q \bar{q}$, with $Q = c, b$
a heavy-quark and $q = u,d,s$ a light-quark.
The properties of heavy mesons and their interactions are expected to be
independent of the heavy-quark spin, which is usually referred to
as heavy-quark spin symmetry (HQSS)~\cite{Isgur:1989vq,Isgur:1989ed}.
The consequences of HQSS for heavy-hadron molecules are important and
have been extensively explored in the literature~\cite{AlFiky:2005jd,Mehen:2011yh,Voloshin:2011qa,Valderrama:2012jv,Nieves:2012tt,Guo:2013sya,Lu:2017dvm,Liu:2018zzu,Liu:2019tjn}.
For S-wave heavy mesons (e.g. the $D$ and $D^*$ charmed mesons) the standard
way to take into account HQSS is to define a superfield $H_Q$ as
\begin{eqnarray}
  H_Q = \frac{1}{\sqrt{2}}\,\left[ P \, {\bf 1} + \vec{\sigma} \cdot
    \vec{P}^* \right] \, ,
\end{eqnarray}
with $P$ and $\vec{P}^*$ the $J^P = 0^-$ and $1^-$ heavy mesons, ${\bf 1}$
the 2 x 2 identity matrix and $\vec{\sigma}$ the Pauli matrices,
where our definition of $H_Q$ corresponds to the non-relativistic
limit of the superfield defined in Ref.~\cite{Falk:1992cx}.
This field has good properties with respect to heavy-quark spin rotation, i.e.
the heavy-quark transformation
$| Q \rangle \to e^{- i \vec{S}_H \cdot \theta} | Q \rangle$
induces the superfield transformation:
\begin{eqnarray}
  H_Q \to e^{- i \vec{S}_H \cdot \theta} H_Q \, ,
\end{eqnarray}
from which it is clear that $H_Q^{\dagger} H_Q \mathcal{O}$ field combinations,
with $\mathcal{O}$ some operator in the form of a 2 x 2 matrix,
will be independent of heavy-quark spin rotations.
With this formalism, the interaction of S-wave heavy mesons with the pion
can be written as
\begin{eqnarray}
  \mathcal{L} = \frac{g_1}{\sqrt{2} f_{\pi}}\,
          {\rm Tr} \left[ H_Q^{\dagger}  H_Q \, \vec{\sigma} \cdot \vec{a} \right]
          \, ,
\end{eqnarray}
where $g_1 = 0.6$ is the axial coupling (a value which is compatible
with $g_1 = 0.59 \pm 0.01 \pm 0.07$ as extracted
from the $D^* \to D \pi$ decay~\cite{Ahmed:2001xc,Anastassov:2001cw}),
$f_{\pi} \simeq 132\,{\rm MeV}$ the pion weak
decay constant and $\vec{a}$ is the (reduced) axial current,
which traditionally only includes the pion
\begin{eqnarray}
  \vec{a} = \vec{\nabla} \pi \, ,
\end{eqnarray}
where we implicitly include the SU(2)-isospin indices in the pion field,
i.e. $\pi = \tau_c \pi^c$ with $c$ an isospin index.

Alternatively, instead of grouping the $P$ and $\vec{P}^*$
into a single superfield with good heavy-quark rotation properties,
we notice that the heavy-quark spin degrees of freedom do not come into play
in the description of heavy-light hadron interactions.
This allows us to write interactions in terms of a fictitious
{\it light-quark subfield}, a heavy field with the quantum numbers of
the light-quark within the heavy meson~\cite{Valderrama:2019sid}.
If we call this effective field $q_L$,
the corresponding Lagrangian will read
\begin{eqnarray}
  \mathcal{L} = \frac{g_1}{\sqrt{2} f_{\pi}}
  q_L^{\dagger} \, \vec{\sigma}_L \cdot \vec{a} \, q_L \, ,
\end{eqnarray}
with $\vec{\sigma}_L$ the spin operators (Pauli matrices)
as applied to the light-quark spin.
When this operator acts on the light-quark degrees of freedom
it can be translated into the corresponding spin operator
acting on the heavy meson field with the rules
\begin{eqnarray}
  \langle P | \vec{\sigma}_L | P \rangle &=& 0 \, , \\
  \langle P | \vec{\sigma}_L | P^* \rangle &=& \vec{\epsilon} \, , \\
  \langle P^* | \vec{\sigma}_L | P^* \rangle &=& \vec{S} \, ,
\end{eqnarray}
with $\vec{\epsilon}$ the polarization vector of the ${P}^*$ heavy meson
and $\vec{S}$ the spin-1 matrices.
From now on we will work in this notation.

The previous light-quark subfield Lagrangian leads to
the non-relativistic potential
\begin{eqnarray}
  V_{\pi}(\vec{q}) &=& - \zeta\,\frac{g_1^2}{2 f_{\pi}^2} \,
  \vec{\tau}_1 \cdot \vec{\tau}_2 \,
  \frac{\vec{\sigma}_{L1} \cdot \vec{q} \, \vec{\sigma}_{L2} \cdot \vec{q}}
       {\vec{q}^2 + m_{\pi}^2} \, , \nonumber \\
       &=& - \zeta\,\frac{g_1^2}{6 f_{\pi}^2} \,
  \vec{\tau}_1 \cdot \vec{\tau}_2 \,
  \frac{\vec{\sigma}_{L1} \cdot \vec{\sigma}_{L2}\,\vec{q}^2}
       {\vec{q}^2 + m_{\pi}^2} \nonumber \\
  && - \zeta\,\frac{g_1^2}{6 f_{\pi}^2} \,
  \vec{\tau}_1 \cdot \vec{\tau}_2 \,
  \frac{\left( 3\,\vec{\sigma}_{L1} \cdot \vec{q} \, \vec{\sigma}_{L2} \cdot \vec{q} -
    \vec{\sigma}_{L1} \cdot \vec{\sigma}_{L2}\,q^2 \right)
  }{\vec{q}^2 + m_{\pi}^2} \, , \nonumber \\
\end{eqnarray}
with $\vec{q}$ the exchanged momentum, $m_{\pi} \simeq 138\,{\rm MeV}$
the pion mass and $\zeta = \pm 1$ a sign which is $+1$($-1$)
for the meson-meson (meson-antimeson) potential
(which comes from the G-parity of the pion).
In the second and third lines we separate the potential
into its S-wave and S-to-D-wave components
(i.e. spin-spin and tensor pieces): owing to the exploratory nature
of the present manuscript, we will be only concerned with the S-wave
components of hadronic molecules and will ignore the D-waves.

Now, to include the axial meson we simply modify
the axial current $\vec{a}$ as follows
\begin{eqnarray}
  \vec{a} = \vec{\nabla} \pi + \lambda_1 m_{a} \vec{a}_1 \, ,
  \label{eq:reduced-axial-current}
\end{eqnarray}
with the isospin indices again implicit, i.e. $a_1 = \tau_c a_{1}^c$, and
$\lambda_1$ a parameter describing how the axial meson mixes
with the pion axial current (which value we will discuss later).
This readily leads to the potential
\begin{eqnarray}
  V_{a_1}(\vec{q}) &=& - \zeta\,\lambda_1^2 \frac{g_1^2 m_{a1}^2}{2 f_{\pi}^2} \,
  \vec{\tau}_1 \cdot \vec{\tau}_2 \,
  \Big[ \frac{\vec{\sigma}_{L1} \cdot \vec{\sigma}_{L2}}
    {\vec{q}^2 + m_{a_1}^2} \nonumber \\
    && \qquad \quad
    + \frac{1}{m_{a_1}^2}\,\frac{\vec{\sigma}_{L1} \cdot \vec{q} \,
      \vec{\sigma}_{L2} \cdot \vec{q}}
    {\vec{q}^2 + m_{a_1}^2} \Big] \, \nonumber \\
  &=& - \zeta\,\lambda_1^2 \frac{g_1^2 m_{a1}^2}{2 f_{\pi}^2} \,
  \vec{\tau}_1 \cdot \vec{\tau}_2 \,
  \frac{\vec{\sigma}_{L1} \cdot \vec{\sigma}_{L2}}
       {\vec{q\,}^2 + m_{a_1}^2} \, ( 1 + \frac{\vec{q\,}^2}{3 m_{a_1}^2})
       + \dots \nonumber \\
       \label{eq:V-a1-p}
\end{eqnarray}
where in the last line we isolate the S-wave component.

Finally we are interested in the r-space expressions of the pion and axial
exchange potentials.
For this we Fourier-transform into r-space, which in the pion case yields
\begin{eqnarray}
  V_{\pi}(\vec{r}) = \zeta\,\frac{g_1^2 m_{\pi}^2}{6 f_{\pi}^2} \,
  \vec{\tau}_1 \cdot \vec{\tau}_2 \,
  \vec{\sigma}_{L1} \cdot \vec{\sigma}_{L2} \,
  \frac{e^{-m_{\pi} r}}{4 \pi r} + \dots \, , \label{eq:V-pi}
\end{eqnarray}
where the dots represent tensor (i.e. S-to-D-wave) and
contact-range (i.e. Dirac-delta) terms (which we also ignore
owing to their short-range nature).
For the axial meson exchange we have instead
\begin{eqnarray}
V_{a_1}(\vec{r}) = -\zeta\, \lambda_1^2 \frac{g_1^2 m_{a1}^2}{3 f_{\pi}^2} \,
  \vec{\tau}_1 \cdot \vec{\tau}_2 \,
  \vec{\sigma}_{L1} \cdot \vec{\sigma}_{L2} \, \frac{e^{-m_{a1} r}}{4 \pi r} + \dots \, ,  \nonumber \\
  \label{eq:V-a1}
\end{eqnarray}
where the dots indicate again contact and tensor terms.

The coupling of the axial meson to the hadrons depends on $\lambda_1$,
which could be deduced from the matrix elements of the axial current $A_{5 \mu}$:
\begin{eqnarray}
  \langle 0 | A_{5 \mu} | \pi \rangle = f_{\pi} q_{\mu} \quad , \quad
  \langle 0 | A_{5 \mu} | a_1 \rangle = f_{a_1} m_{a_1}\,\epsilon_{\mu} \, ,
\end{eqnarray}
with $q_{\mu}$ the momentum of the pion, $f_{\pi}$ and $f_{a_1}$ the weak decay
constants of the pion and axial meson, $m_{{a}_1}$ the mass of the axial meson
and $\epsilon_{\mu}$ its polarization vector.
From Eq.~(\ref{eq:axial-current}) we arrive at the identification
\begin{eqnarray}
  \lambda_1 = \frac{f_{a_1}}{f_{\pi}} \, ,
\end{eqnarray}
but $f_{a_1}$ is not particularly well-known.
Different estimations exist, of which a few worth noticing are:
\begin{itemize}
\item[(i)] The  Weinberg sum rules~\cite{Weinberg:1966fm} or
  the Kawarabayashi-Suzuki-Riazuddin-Fayyazuddin (KSFR)
  relations~\cite{Kawarabayashi:1966kd,Riazuddin:1966sw},
  both of which lead to $m_{a_1} = \sqrt{2}\,m_{\rho} = 1.09\,{\rm GeV}$,
  $f_{a_1} = f_{\pi}$ and $\lambda_1 = 1$.
\item[(ii)] The $\tau \to 3\pi \, \nu_{\tau}$ decay involves the axial meson
  as an intermediate state, and have been used in the past to determine
  $f_{a_1}$:
  \begin{itemize}
  \item[(ii.a)] Three decades ago Ref.~\cite{Isgur:1988vm} obtained
    \begin{eqnarray}
      m_{a_1} f_{a_1} = (0.25 \pm 0.02)\,{\rm GeV}^2 \, ,
    \end{eqnarray}
    for $m_{a_1} = 1.22\,{\rm GeV}$, which translates into
    $\lambda_1 = 1.55 \pm 0.12$.
    Later Ref.~\cite{Wingate:1995hy} made the observation that $m_{a_1} f_{a_1}$
    shows a simple dependence on the $\tau \to 3\pi \, \nu_{\tau}$ branching
    ratio, from which it updated the previous value to
    $m_{a_1} f_{a_1} = (0.254 \pm 0.20)\,{\rm GeV}^2$,
    yielding $\lambda_1 = 1.58 \pm 0.12$.
  \item[(ii.b)]
    Ref.~\cite{Coon:1996wj} noticed a result from Ref.~\cite{Pham:1991ex},
    which contains a phenomenological relation between $m_{a_1} f_{a_1}$ and
    the relative branching ratios for the $\tau \to 2\pi \nu_{\tau}$ and
    $\tau \to 3\pi \nu_{\tau}$ decay. This led the authors
    of Ref.~\cite{Coon:1996wj} to the estimation~\footnote{We notice that
      Ref.~\cite{Coon:1996wj} uses the $f_{\pi} \sim 93\,{\rm MeV}$
      normalization for the decay constants, i.e. a $\sqrt{2}$ factor
      smaller than ours. Thus we have adapted their results to
      our normalization.
    }
    \begin{eqnarray}
    m_{a_1} f_{a_1} = 0.33\,{\rm GeV}^2 \, ,
    \end{eqnarray}
    which is equivalent to $\lambda_1 = 2.05$.
  \item[(ii.c)]
    Chiral Lagrangian analyses of the $\tau \to 3\pi \nu_{\tau}$
    decay~\cite{Dumm:2009va,Nugent:2013hxa}, usually yield
    $\lambda_1 \sim 1.4-1.5$ but with $m_{a_1} \sim 1.1\,{\rm GeV}$,
    which is somewhat light.
  \end{itemize}
\item[(iii)] The lattice QCD calculation of Ref.~\cite{Wingate:1995hy} gives
  \begin{eqnarray}
    m_{a_1} f_{a_1} = (0.30 \pm 0.02)\,{\rm GeV}^2 \, ,
  \end{eqnarray}
  and $m_{a_1} = 1.25 \pm 0.08 \,{\rm GeV}$, from which we extract
  $\lambda_1 = 1.82 \pm 0.08 \pm 0.12$ where the first and second error refer
  to the uncertainties in $m_{a_1} f_{a_1}$ and $m_{a_1}$, respectively.
\item[(iv)] Ref.~\cite{Yang:2007zt} uses QCD sum rules to obtain
  \begin{eqnarray}
    f_{a_1} = (238 \pm 10)\,{\rm MeV} \, ,
  \end{eqnarray}
  that is, $\lambda_1 = 1.80 \pm 0.08$.
\end{itemize}
From the previous, it is apparent that
the uncertainties in $\lambda_1$ are large.
But we can reduce its spread if we concentrate on the determinations of
$\lambda_1$ for which $m_{a_1}$ is close to its value
in the Review of Particle Physics (RPP)~\cite{Zyla:2020zbs},
i.e. $m_{a_1} = 1.23\,{\rm GeV}$ (which is also the value
we will adopt for the mass of the axial meson).
In this case we end up with the $\lambda_1 \sim (1.55-2.05)$ window,
which we will approximate by $\lambda_1 = 1.8 \pm 0.3$.
This is the central value and uncertainty we will use from now on.

At this point it is interesting to compare the strengths of the resulting
Yukawa-like piece of the previous potentials
\begin{eqnarray}
  V_Y(\vec{r}) = \pm \frac{g_Y^2}{4 \pi} \mathcal{O}_I \mathcal{O}_S\,
  \frac{e^{-m r}}{r} \, , \label{eq:Yukawa-like}
\end{eqnarray}
where $g_Y$ is an effective Yukawa-like coupling,
$\mathcal{O}_I = 1$ or $\vec{\tau}_1 \cdot \vec{\tau}_2$ and
$\mathcal{O}_S = 1$ or $\vec{\sigma}_{L1} \cdot {\sigma}_{L2}$
the usual isospin and spin operators, while $m$ is the mass of
the exchanged meson.
For the pion and axial meson exchange potentials we have that the strength
of the effective Yukawas are
\begin{eqnarray}
  \frac{g_{\,\rm Y(\pi)}^2}{4 \pi} \simeq 6.6 \cdot 10^{-2} \quad \mbox{and} \quad
  \frac{g_{\,\rm Y(a_1)}^2}{4 \pi} \simeq 2.0-3.5 \, ,
\end{eqnarray}
which gives an idea of the relative strength of axial meson exchange with
respect to the pion.
Provided it is attractive, the condition for this effective Yukawa-like
potential to bind is
\begin{eqnarray}
  \frac{2 \mu}{m}\,\frac{g_{Y}^2}{4 \pi} \left|
  \langle \mathcal{O}_I \rangle \langle \mathcal{O}_S \rangle
  \right| \geq 1.68 \, ,
\end{eqnarray}
with $\mu$ the reduced mass of the two hadron system.
If we consider the $I^G(J^{PC}) = 1^{+}(1^{+-})$ $D^{*} \bar{D}$ system, which
is the usual molecular interpretation of the $Z_c(3900)$,
the potential is indeed attractive and the previous
condition is fulfilled for $\lambda_1 \geq 1.1$
(if $m_{a_1} = 1.23\,{\rm GeV}$).

\section{Scalar and vector meson exchange}
\label{sec:vector}

Besides the pion and the axial mesons, usually the other important exchanged
light-mesons in the OBE model are the scalar $\sigma$ and the vector mesons
$\rho$ and $\omega$.
In the following lines we will discuss the potentials they generate.

\subsection{Scalar meson}

For the scalar meson we write a Lagrangian of the type
\begin{eqnarray}
  \mathcal{L}_S &=&
  g_{\sigma 1} \, {\rm Tr} \left[ H_Q^{\dagger} H_Q \right]\,\sigma \\
  &=&
  g_{\sigma 1} \, q_L^{\dagger} \sigma q_L \, ,
\end{eqnarray}
depending on the notation (superfield/subfield in first/second line),
with $g_{\sigma 1}$ the coupling of the scalar meson to
the charmed hadrons.
From this Lagrangian we derive the potential
\begin{eqnarray}
  V_{\sigma}(\vec{q}) &=& - \frac{g_{\sigma 1}^2}{\vec{q}^2 + m_\sigma^2} \, ,
\end{eqnarray}
which is attractive and where $m_{\sigma}$ is the scalar meson mass.
Finally, if we Fourier-transform into coordinate space we will arrive at
\begin{eqnarray}
  V_{\sigma}(\vec{r}) &=& - g_{\sigma 1}^2\,\frac{e^{-m_{\sigma} r}}{4 \pi r} \, .
  \label{eq:V-sigma}
\end{eqnarray}

The parameters in this potential are the coupling $g_{\sigma 1}$ and
the mass $m_{\sigma}$.
For the coupling we will rely on the linear sigma model
(L${\sigma}$M)~\cite{GellMann:1960np}, which we briefly review here
as it will prove useful for the discussion on the $Z_{cs}(3985)$ later.
The L${\sigma}$M is a phenomenological model in which originally
we have a massless nucleon field that couples to a combination of
four boson fields, i.e. this model contains a nucleon interaction term of
the type
\begin{eqnarray}
  \mathcal{L}_{\rm int}^{\rm N \sigma L} = g\,\bar{\psi}_N 
    ( \phi_0 + i\, \gamma_5 \, \vec{\tau} \cdot \vec{\phi} ) \psi_N \, ,
\end{eqnarray}
where $\psi_N$ is the relativistic nucleon field and $g$ a coupling constant.
By means of spontaneous symmetry breaking we end up with three massless bosons
$\vec{\phi}$, which might be interpreted as pions, while the isospin scalar
$\phi_0$ acquires a vacuum expectation value
($\langle \phi_0 \rangle = f_{\pi} / \sqrt{2}$)
which also provides the nucleons with mass.
The $\sigma$ field is defined as a perturbation of the $\phi_0$ field
around its vacuum expectation value ($\phi_0 = f_{\pi}/\sqrt{2} + \sigma$).
This model provides a relation between $f_{\pi}$, the nucleon mass
$M_N \simeq 940\,{\rm MeV}$ and the couplings of the scalar mesons
and the pions to the nucleon,
where $g = g_{\sigma NN} = g_{\pi NN} = \sqrt{2} M_N / f_{\pi} = 10.2$.

Nowadays we know that the pion coupling is derivative (as required by chiral
symmetry), yet if we are considering one-pion exchange only this derivative
coupling can be matched to a non-derivative one as in both cases
the same potential is obtained.
In this case the pion coupling is given by
$g_{\pi NN} = g_A\,\sqrt{2} M_N / f_{\pi}$ with $g_A = 1.26$,
which means that the linear sigma model is off by about a $26\%$
(or a $30\%$ once we take into account the Goldberger-Treiman discrepancy).
Thus this is the expected uncertainty that we should have for $g_{\sigma NN}$.
For comparison purposes, the L$\sigma$M gives
$g_{\sigma NN}^2 / 4 \pi = 8.3$ while the OBE model of nuclear
forces~\cite{Machleidt:1987hj,Machleidt:1989tm} prefers slightly
larger values $g_{\sigma NN}^2 / 4 \pi = 8.5-8.9$ (which are still
compatible with the L${\sigma}$M).
For the charmed mesons, which contain only one light-quark, we will assume
the quark model relation
$g_{\sigma 1} = g_{\sigma NN} /3 \simeq 3.4$~\cite{Riska:2000gd}
(though we notice that Ref.~\cite{Riska:2000gd} advocates a slightly
larger coupling of $g_{\sigma qq} = \sqrt{2}\,m_q^{\rm con} / f_{\pi} \simeq 3.6$,
with $m_q^{\rm con}$ the constituent $q=u,d$ quark mass).

For the mass of the sigma the OBE model of nuclear forces uses
$m_{\sigma} = 550\,{\rm MeV}$, but it is also common to find
$m_{\sigma} = 600\,{\rm MeV}$ in a few recent implementations of
the OBE model for hadronic molecules~\cite{Liu:2008tn,Sun:2011uh,Liu:2018bkx,Liu:2019stu,Liu:2019zvb}.
Nowadays the RPP designation of the $\sigma$ is $f_0(500)$ and
the mass is in the $400-550\,{\rm MeV}$ range~\cite{Zyla:2020zbs}.
However this does not necessarily imply that the mass of the $f_0(500)$ pole
should be used for the scalar meson exchange, owing in part to its large
width and in part to its relation with correlated two-pion exchange,
as has been extensively discussed~\cite{Holzenkamp:1989tq,Reuber:1995vc,Oset:2000gn,CalleCordon:2009ps}.
Direct fits of $g_{\sigma NN}$ and $m_{\sigma}$ can also lead to more
than one solution, though they are usually compatible
with the RPP mass range of the sigma and with the expected $30\%$
uncertainty for the coupling in the L${\sigma}$M.
For instance, a renormalized OBE fit to NN data~\cite{CalleCordon:2008eu}
leads to two solutions, one with $m_{\sigma} = 477\,{\rm MeV}$, $g_{\sigma NN} = 8.76$ and another with $m_{\sigma} = 556\,{\rm MeV}$, $g_{\sigma NN} = 13.04$.
What we will do then is to investigate binding as a function of
the $\sigma$ mass.

\subsection{Vector mesons}

The interaction of the vector mesons with hadrons is analogous to
that of the photons and it can be expanded
in a multipole expansion.
For the S-wave charmed mesons the spin of the light-quark degree of freedom
is $S_L = \tfrac{1}{2}$, which admits an electric charge ($E0$) and
magnetic dipole ($M1$) moment, from which the Lagrangian reads
\begin{eqnarray}
  \mathcal{L}_V &=& \mathcal{L}_{E0} + \mathcal{L}_{M1} \nonumber \\
  &=& g_{V1} {\rm Tr} \left[ H_Q^{\dagger} H_Q \right] V^0 \nonumber \\
  &+& \frac{f_{V1}}{2 M}\,\epsilon_{ijk}\,{\rm Tr}
  \left[ H_Q^{\dagger} H_Q \sigma_i \right] \partial_j V_k \, \\
  &=& q_L^{\dagger} \left[ g_{V1} V^0  + \frac{f_{V1}}{2 M}\,
  \epsilon_{ijk} \sigma_{Li} \partial_j V_k \right] q_L
  \, ,
\end{eqnarray}
depending on the notation (superfield or subfield),
where $g_{V1}$ and $f_{V1}$ are the electric- and magnetic-type couplings
with the S-wave charmed mesons, $M$ is a mass scale (it will prove
convenient to choose this mass scale equal to the nucleon mass,
i.e. $M = M_N$) and $V^{\mu} = (V^0, \vec{V})$ the vector meson field.
For notational convenience we have momentarily ignored the isospin factors.
From this the vector-meson exchange potentials are also expressible as a sum
of multipole components
\begin{eqnarray}
  V_V(\vec{q}) &=& V_{E0}(\vec{q}) + V_{M1}(\vec{q}) \, ,
\end{eqnarray}
which read
\begin{eqnarray}
  V_{E0}(\vec{q}) &=& + \frac{g_{V1}^2}{\vec{q}^2 + m_V^2} \, , \\
  V_{M1}(\vec{q}) &=& - \frac{f_{V1}^2}{4 M^2}
  \frac{(\vec{\sigma}_{L1} \times \vec{q}) \cdot (\vec{\sigma}_{L2} \times \vec{q})}{\vec{q}^2 + m_V^2} \nonumber \\
  &=& - \frac{f_{V1}^2}{6 M^2}\,\vec{\sigma}_{L1} \cdot \vec{\sigma}_{L2} \,
  \frac{\vec{q}^2}{\vec{q}^2 + m_V^2} + \dots \, ,
\end{eqnarray}
with $m_V$ the vector meson mass and where the second line of
the M1 contribution to the potential isolates
its S-wave component.
After Fourier-transforming into coordinate space we end up with
\begin{eqnarray}
  V_V(\vec{r}) &=& \left[ g_{V1}^2 +
  f_{V1}^2\frac{m_V^2}{6 M^2}\,\vec{\sigma}_{L1} \cdot \vec{\sigma}_{L2} \right] \, \frac{e^{-m_V r}}{4 \pi r} + \dots \, , \nonumber \\
\end{eqnarray}
where the dots indicate contact-range and tensor terms, which we are ignoring.
If we particularize for the $\rho$ meson, we will have to include isospin
factors
\begin{eqnarray}
  V_{\rho}(\vec{r}) &=& \vec{\tau}_1 \cdot \vec{\tau}_2\,\left[ g_{\rho 1}^2 +
    f_{\rho 1}^2\frac{m_\rho^2}{6 M^2}\,\vec{\sigma}_{L1} \cdot \vec{\sigma}_{L2} \right] \, \frac{e^{-m_{\rho} r}}{4 \pi r} + \dots \, . \nonumber \\
  \label{eq:V-rho}
\end{eqnarray}
For the $\omega$ no isospin factor is required, but there is a sign coming
from the negative G-parity of this meson
\begin{eqnarray}
  V_{\omega}(\vec{r}) &=& \zeta\,\left[ g_{\omega 1}^2 +
    f_{\omega 1}^2\frac{m_\omega^2}{6 M^2}\,\vec{\sigma}_{L1} \cdot \vec{\sigma}_{L2} \right] \, \frac{e^{-m_{\omega} r}}{4 \pi r} + \dots \, , \nonumber \\
  \label{eq:V-omega}
\end{eqnarray}
where, as usual, $\zeta = +1$ ($-1)$ for the meson-meson (meson-antimeson)
potential.

The determination of the couplings with the vector mesons follows
the same pattern we have used for the axial mesons.
The neutral vector mesons, the $\omega$ and the $\rho^3$ (where $^3$ refers to
the isospin index, i.e. the neutral $\rho$), have the same quantum
numbers as the photon and thus can mix with the electromagnetic current.
It is convenient to write down the mixing in the form
\begin{eqnarray}
  \rho^3_{\mu} &\to& \rho^3_{\mu} + \lambda_{\rho}\,\frac{e}{g}\,A_{\mu} \, ,
  \label{eq:rho-em-mixing} \\
  \omega_{\mu} &\to& \omega_{\mu} + \lambda_{\omega}\,\frac{e}{g}\,A_{\mu} \, ,
  \label{eq:omega-em-mixing} 
\end{eqnarray}
with $e$ the electric charge of the proton and
$g = m_V / 2 f_{\pi} \simeq 2.9$ the universal vector meson coupling
constant.
These two substitution rules effectively encapsulate Sakurai's universality and
vector meson dominance~\cite{Sakurai:1960ju,Kawarabayashi:1966kd,Riazuddin:1966sw}.

The proportionality constants can be determined from matching with
the electromagnetic Lagrangian of the light-quark components of the hadrons.
To illustrate this idea,
we can apply the substitution rules to the E0 piece of the Lagrangian
describing the interaction of the neutral vector mesons with the charmed
antimesons, i.e.
\begin{eqnarray}
  \mathcal{L}_{E0} &=&
          {\rm Tr} \left[ H_{\bar c}^{\dagger} \left( g_{\rho 1}\,\tau_3 \rho^3_0 +
            g_{\omega 1}\,\omega_0 \right) H_{\bar c} \right] \, ,
\end{eqnarray}
where we have chosen the antimesons because they contain light-quarks.
After applying Eqs.~(\ref{eq:rho-em-mixing}) and (\ref{eq:omega-em-mixing}),
we end up with
\begin{eqnarray}
  \mathcal{L}_{E0}^{\rm e.m. (L)} &=&
  e\,  {\rm Tr} \left[ H_{\bar c}^{\dagger}
    \left( \frac{g_{\rho 1}}{g} \lambda_{\rho}\,\tau_3 +
    \frac{g_{\omega 1}}{g} \lambda_{\omega} \right) H_{\bar c}
    \right] A_0  \nonumber \\
  &=&
  e\,  {\rm Tr} \left[ H_{\bar{c}}^{\dagger}
    \left( \lambda_{\rho}\,\tau_3 + \lambda_{\omega} \right) H_{\bar c}
    \right] A_0 \, .
\end{eqnarray}
where in the second line we have used that $g_{\rho 1} = g_{\omega_1} = g$.
This is to be matched with the contribution of the light-quarks to
the E0 electromagnetic Lagrangian
\begin{eqnarray}
  \mathcal{L}_{E0}^{\rm e.m.} &=&
  e\,  {\rm Tr} \left[ H_{\bar Q}^{\dagger}\,
    \left( Q_H + Q_L \right) H_{\bar Q} \right] A^0 \, .
\end{eqnarray}
where $Q_H$ and $Q_L$ are the electric charges of the heavy-antiquark and
light-quarks in the isospin basis of the superfield $H_{\bar Q}$,
of which only $Q_L$ is relevant for matching purposes
\begin{eqnarray}
  Q_L &=&
  \begin{pmatrix}
    \frac{2}{3} & 0 \\
    0 & -\frac{1}{3}
  \end{pmatrix} \, .
\end{eqnarray}
which implies that $\lambda_{\rho} = 1/2$ and $\lambda_{\omega} = 1/6$.
Alternatively we could have determined $\lambda_{\rho}$ and $\lambda_{\omega}$
from the nucleon couplings to the vector mesons ($g_{\rho} = g$,
$g_{\omega} = 3 g$) and their electric charges,
leading to the same result.

Given $\lambda_{\rho}$ and $\lambda_{\omega}$ and repeating the same steps
but now for the M1 part of the Lagrangian, we can readily infer
the magnetic-type coupling $f_{V1}$ of the charmed antimesons
with the vector mesons, which turn out to be
\begin{eqnarray}
  f_{V1} = g \kappa_{V1} \quad \mbox{with} \quad 
  \kappa_{V1} = \frac{3}{2}\left(\frac{2 M}{e}\right) \mu_L(D^{*0}) \, ,
\end{eqnarray}
where $\mu_L(D^{*0})$ refers to the light-quark contribution to
the magnetic moment of the $D^{*0}$ charmed antimeson,
which in the heavy-quark limit will coincide
with the total magnetic moment of
the heavy meson.
From the quark model we expect this magnetic moment to be given by
the u-quark, i.e.
\begin{eqnarray}
  \left(\frac{2 M}{e}\right) \mu_L(D^{*0}) =
  \left(\frac{2 M_N}{e}\right) \mu_u \simeq 1.85  \, ,
\end{eqnarray}
where we have taken $M = M_N \simeq 940\,{\rm MeV}$ (i.e. the nucleon mass)
as to express the magnetic moments in units of nuclear magnetons.

The outcome is $g_{V1} = 2.9$ and $\kappa_{V1} = 2.8$,
which are the values we will use here.
Besides this determination, the vector meson dominance model of
Ref.~\cite{Casalbuoni:1992dx} leads to $g_{V1} = 2.6$ and $\kappa_{V1} = 2.3$
(as explained in more detail in Ref.~\cite{Liu:2019stu}),
i.e. compatible with our estimates within the $20\%$ level.
For the particular case of the E0 coupling, there is a lattice QCD calculation
for the heavy mesons~\cite{Detmold:2012ge}
yielding $g_{V1} = 2.6 \pm 0.1 \pm 0.4$ in the heavy quark limit
(i.e. compatible within errors with $g_{V1} = 2.9$).

\section{Description of the $X(3872)$ and $Z_c(3900/4020)$}
\label{sec:Zc}

Now we will consider the $X(3872)$, $Z_c(3900)$ and $Z_c(4020)$
from the OBE model perspective.
The problem we want to address is:
can they be described together with the same set of parameters?
We will find that
\begin{enumerate}
\item[(i)] the axial meson indeed favors the compatible description
  of the $X$ and $Z_c$ resonances,
\item[(ii)] yet the effect of axial mesons depends on the choice of a mass
  for the scalar meson in the OBE model.
\end{enumerate}
In general, lighter scalar meson masses will diminish the impact of
axial meson exchange and eventually even vector meson exchange,
leading to the binding of both the $X$ and $Z_c$ for
$m_{\sigma} \to 400\,{\rm MeV}$.
This is not necessarily a desired feature, as the $Z_c$ in the molecular picture
is not necessarily a bound state but more probably a virtual state
or a resonance~\cite{Albaladejo:2015lob,Yang:2020nrt}.
That is, we expect the strength of the charmed meson-antimeson
potential to be short of binding for the $Z_c$ and $Z_c^*$.
However, as the mass of the scalar meson increases and reaches the standard
values traditionally used in the OBE model, $m_{\sigma} \sim 500-600\,{\rm MeV}$,
the importance of the axial meson becomes clearer, where the $a_1$ meson
might be the difference between a virtual state close to threshold or not.

\subsection{Molecular degrees of freedom:
  which interpretation to choose?}
\label{subsec:channels}

The nature of the $X(3872)$ and the $Z_c(3900/4020)$ is still an open problem.
Here we will assume that they are molecular, that is, that they are
two-meson states.
This requires to identify two-meson thresholds close to the masses of
the $X(3872)$ and $Z_c(3900/4020)$ states and compatible
with their quantum numbers.
The most obvious candidates are the different charmed meson-antimeson
combinations, e.g. the $D^*\bar{D}$ for the $X(3872)$ and $Z_c(3900)$
and the $D^* \bar{D}^*$ for the $Z_c(4020)$.
This will be the choice we will make in the present work.

However, this is not the only possibility.
For instance, some interpretations assume that the $X(3872)$
contains a $J/\psi \omega$ component~\cite{Swanson:2003tb},
to which we may include $J/\psi \rho$ if isospin breaking effects
are explicitly considered.
For the $Z_c(3900/4020)$, if we assume that their quantum numbers are indeed
$I^G (J^{PC}) = 1^+ (1^{+-})$, they could also contain a $\eta_c \rho$
component~\cite{Aceti:2014uea} (though this channel is located further
from the $Z_c$'s than the charmed meson-antimeson components).
If we extend this argument to the $Z_{cs}(3985)$, besides the charmed
meson-antimeson $D_s^*\bar{D}$ and $D_s\bar{D}^*$ components, we could
also add $\eta_c K^*$ (or even $J/\psi\,K$)~\cite{Ikeno:2020csu}.

Though these degrees of freedom have been explicitly considered in other works, 
we will not include them.
The reason is their expected relative strength and range when compared to
the other meson exchanges considered here.
To illustrate this idea, we might consider the $\eta_c \rho$ component
in the $Z_c(3900)$, for which the $\eta_c \rho \to D^* \bar{D}$
transition potential is mediated by charm vector meson exchange.
The form of this potential can in principle be deduced in a way analogous
to the vector meson exchange potential, leading to
\begin{eqnarray}
  \langle \eta_c \rho | V_{D^*}(\vec{q}) | D^{*} \bar{D} (Z_c)
  \rangle  = \frac{h_1 h_2}{\mu_{D^*}^2 + {\vec{q}\,}^2} \, ,
  \label{eq:V-mDD-p}
\end{eqnarray}
where for simplicity we have only considered the E0 component;
``$(Z_c)$'' indicates that we are already projecting into the $Z_c(3900)$
channel, $h_{1(2)}$ are the relevant coupling constants in the vertices $1(2)$,
$\mu_{D}^*$ is the effective mass of the exchanged $D^*$ meson, which
is somewhat lighter than its physical mass owing to the fact that
the $D^*$ meson has a non-trivial zero-th component of
its 4-momentum (check Sect.~\ref{subsec:mass-gaps}
for a more detailed explanation).
This potential is relatively short-ranged, but (light) vector meson
exchange cancels out in the $Z_c$ channel, meaning
that $D^*$ exchange could be more important
than expected.

But this conclusion still depends on the absence of other light-meson
exchange contributions that could mask $D^*$ exchange.
In this work we have at least two of these contributions: scalar and axial
meson exchange.
If we compare the expected strength of $a_1$ and $D^*$ exchanges at
low momenta (i.e.. the potentials of Eqs.~(\ref{eq:V-a1-p})
and (\ref{eq:V-mDD-p}) at $| {\vec{q}\,}^2 | \to 0$),
their ratio will be
\begin{eqnarray}
  \lim_{{\vec{q}\,}^2 \to 0}\left| \frac{V_{D^*}(\vec{q}; Z_c)}{V_{a_1}(\vec{q}; Z_c)} \right| = \frac{h_1 h_2}{\mu_{D^*}^2}\,\frac{2 f_{\pi}^2}{3 \lambda_1^2 g_1^2}
  \simeq 0.04
  \, .
\end{eqnarray}
where following Ref.~\cite{Ikeno:2020csu} we have taken
$h_1 = \sqrt{2} g_V (m_D + m_{\eta_c})/\sqrt{2 m_D} \sqrt{2 m_{\eta_c}} \simeq 1.45 g_V$ and $h_2 = g_V$, while for the effective mass we have used
$\mu_{D^*} \simeq \sqrt{m_{D^*}^2 - (m_{\eta_c}^2 - m_D^2)} \simeq 1.67\,{\rm GeV}$
(check Sect.~\ref{subsec:mass-gaps}).
This preliminary comparison indicates that the contribution from $D^*$ exchange 
can be probably neglected if we have already included axial meson exchange.
However if neither scalar nor axial meson exchanges are included or are
expected to be weaker than here, it will make sense to include
$D^*$ exchange and the $\eta_c \rho$ channel.
It is also worth mentioning that there is another factor reducing
the potential importance of the aforementioned $\eta_c \rho$ channel:
it is located about $200\,{\rm MeV}$ below the $Z_c(3900)$.
As a consequence, iterations of the $D^*$-exchange potential will be further
suppressed owing to this mass gap.
Yet, a caveat is in place: the previous argument does not take into account
the effect of form factors (see Sect.~\ref{subsec:ff}),
which could be very different in the $a_1$ and $D^*$ exchange cases,
or all the possible channels or meson exchanges involved.
These effects could increase the importance of channels
other than charmed meson-antimeson.

\subsection{General structure of the potential}

Before considering the light-meson exchange potential in detail,
we will review the general structure of the S-wave potential.
For the $D^{(*)} \bar{D}^{(*)}$ system there are two relevant symmetries
--- SU(2)-isospin and HQSS --- from which we decompose
the potential into
\begin{eqnarray}
  V = (V_a + \tau W_a) + (V_b + \tau W_b)\,
  \vec{\sigma}_{L1} \cdot \vec{\sigma}_{L2} \, ,
  \label{eq:V-light-spin}
\end{eqnarray}
with $\tau = \vec{\tau}_1 \cdot \vec{\tau}_2$.
In this notation, the $X$ and $Z$ potentials read
\begin{eqnarray}
  V_X &=& (V_a - 3 W_a) + (V_b -3 W_b) \, , \\
  V_Z &=& (V_a + W_a) - (V_b + W_b) \, .
\end{eqnarray}
However, it will be more useful to define the isoscalar and isovector
contributions to the potential as follows
\begin{eqnarray}
  V^{(0)}_a &=& V_a - 3 W_a \, , \\
  V^{(0)}_b &=& V_b - 3 W_b \, , \\  
  \nonumber \\
  V^{(1)}_a &=& V_a + W_a \, , \\
  V^{(1)}_b &=& V_b + W_b \, , \label{eq:V1b} 
\end{eqnarray}
from which the general structure of the potential is
the one shown in Table~\ref{tab:potential-XZ}.
In the following lines we will explain what are the contributions of
each light-meson to the potential, yet we can advance that
\begin{itemize}
\item[(i)] $V^{(I)}_a$ receives contributions from $\sigma$ and E0 $\rho$/$\omega$ and is attractive for $D^{(*)} \bar{D}^{(*)}$ molecules.
\item[(ii)] $V^{(I)}_b$ receives contributions from the pion, the axial meson
  and M1 $\rho$/$\omega$ exchange:
  \begin{itemize}
  \item[(ii.a)] $V^{(0)}_b$ is dominated by M1 vector meson exchange and
    its sign is negative, making the $X(3872)$ the most attractive
    isoscalar molecular configuration.
  \item[(ii.b)] $V^{(1)}_b$ is dominated by axial meson exchange and its sign
    is positive, implying that the $Z_c(3900)$ and $Z_c(4020)$ are among
    the most attractive isovector molecular configurations.
  \item[(ii.c)] The most attractive isovector configuration should be
    the $I^G(J^{PC}) = 1^{-}(0^{++})$ $D^{*} \bar{D}^{*}$ molecule,
    though no molecular state has been found yet with these
    quantum numbers.
  \end{itemize}
  \end{itemize}
Finally, the potentials for the $Z_c(3900)$ and $Z_c(4020)$ are identical,
which explains the evident observation that
they come in pairs~\cite{Guo:2013sya}.
For this reason from now on we will ignore the $Z_c(4020)$ and concentrate
in the $Z_c(3900)$, as results in the later automatically
apply to the former.

\begin{table*}[!ttt]
\begin{tabular}{|lclc|lclc|}
\hline\hline
System  & $I^G(J^{PC})$ & Potential & Candidate &
System  & $I^G(J^{PC})$ & Potential & Candidate 
\\
\hline
$D\bar{D}$ & $0^{+}(0^{++})$ & $V_a^{(0)}$ & - &
$D\bar{D}$ & $1^{-}(0^{++})$ & $V_a^{(1)}$ & - \\
$D^*\bar{D}$ & $0^{-}(1^{+-})$ & $V_a^{(0)} - V_b^{(0)}$ & - &
$D^*\bar{D}$ & $1^{+}(1^{+-})$ & $V_a^{(1)} - V_b^{(1)}$ & $Z_c(3900)$ \\
$D^*\bar{D}$ & $0^{+}(1^{++})$ & $V_a^{(0)} + V_b^{(0)}$ & $X(3872)$ &
$D^*\bar{D}$ & $1^{-}(1^{++})$ & $V_a^{(1)} + V_b^{(1)}$ & - \\
$D^*\bar{D}^*$ & $0^{+}(0^{++})$ & $V_a^{(0)} - 2\, V_b^{(0)}$ & - &
$D^*\bar{D}^*$ & $1^{-}(0^{++})$ & $V_a^{(1)} - 2\, V_b^{(1)}$ & - \\
$D^*\bar{D}^*$ & $0^{-}(1^{+-})$ & $V_a^{(0)} - V_b^{(0)}$ & - &
$D^*\bar{D}^*$ & $1^{+}(1^{+-})$ & $V_a^{(1)} - V_b^{(1)}$ & $Z_c(4020)$ \\
$D^*\bar{D}^*$ & $0^{+}(2^{++})$ & $V_a^{(0)} +\, V_b^{(0)}$ & - &
$D^*\bar{D}^*$ & $1^{-}(2^{++})$ & $V_a^{(1)} +\, V_b^{(1)}$ & - \\
  \hline\hline
\end{tabular}
\caption{SU(2)-isospin and HQSS structure of the S-wave potential
  in the heavy meson-antimeson molecules.
  ``System'' indicates the specific charmed meson-antimeson molecule,
  $I(J^{PC})$ its quantum numbers, ``Potential'' the potential and
  ``Candidate'' refers to known experimental resonances that might
  be explained by the specific configuration considered.
  $V_a^{(I)}$ and $V_b^{(I)}$ are the central and spin-spin pieces of
  the potential, with $I = 0,1$ referring to the isospin.
  From light-meson exchanges we expect $V_a^{(0)} < 0$ and $V_b^{(0)} < 0$
  in the isoscalar sector, which makes the $1^{++}$ and $2^{++}$
  the most promising configurations for binding (in the absences of
  other binding factors, e.g. coupled channels, nearby charmonia,
  etc.).
  For the isovector sector we expect $V_a^{(1)} < 0$ and $V_b^{(1)} > 0$,
  from which the $0^{++}$ and $1^{+-}$ configurations are
  the most promising. However $V_b^{(1)}$ is really weak,
  making this conclusion contingent on other factors
  (e.g. isospin breaking in vector meson exchange).
}
\label{tab:potential-XZ}
\end{table*}

\subsection{The OBE potential}

In the OBE model the $H_c H_c'$ and $H_c \bar{H}_c'$ potentials
(where $H_c$, $H_c'$ represent the S-wave charmed mesons)
can be written as the sum of each light-meson contribution
\begin{eqnarray}
  V_{\rm OBE} = V_{\pi}^{(\zeta)} + V_{\sigma} + V_{\rho} + V_{\omega}^{(\zeta)} +
  V_{a_1}^{(\zeta)} \, , \label{eq:OBE-potential}
\end{eqnarray}
where the individual contributions have been already discussed
in this manuscript (for the approximation in which the hadrons
are point-like):
\begin{itemize}
\item[(i)] $V_{\pi}^{(\zeta)}$ and $V_{a1}^{(\zeta)}$
in Eqs.~(\ref{eq:V-pi}) and (\ref{eq:V-a1}),
\item[(ii)] $V_{\sigma}$ in Eq.~(\ref{eq:V-sigma}),
\item[(iii)] $V_{\rho}$ and $V^{(\zeta)}_{\omega}$
in Eqs.~(\ref{eq:V-rho}) and (\ref{eq:V-omega}).
\end{itemize}
We have included the superscript ${}^{(\zeta)}$ as a reminder that
the contributions stemming from exchange of negative G-parity
light mesons ($\pi$, $\omega$, $a_1$) change sign depending
on whether we are considering the meson-meson ($\zeta = +1$)
or meson-antimeson ($\zeta = -1$) systems.
These signs have been already included in the definition of
the potential contributions, i.e. in Eqs.~(\ref{eq:V-pi}), (\ref{eq:V-a1})
and (\ref{eq:V-omega}).
For convenience we review our choice of couplings
in Table \ref{tab:couplings}.

Here we consider only the S-wave component of the light-meson exchange
potential, i.e. we ignore the tensor (S-to-D-wave) components.
This choice allows a simpler analysis of the factors involved in binding.

{Finally, we will assume the OBE model to be a fairly complete description
  of the charmed meson-antimeson potential.
  Even though there are shorter-range components of the potential, e.g.
  the previously discussed transition potentials into
  the charmonium -- (light) meson channels or the vector
  charmonium exchange potentials considered
  in Refs.~\cite{Aceti:2014kja,Aceti:2014uea,He:2015mja,Dong:2021juy},
  if we follow the arguments of Sect~\ref{subsec:channels}.
  these pieces of the potential should be suppressed with respect to (light)
  meson exchange.
  Nonetheless they could be easily included as a contact-range potential
  following the general structure of Table \ref{tab:potential-XZ}.
  This type of improved OBE with contacts has been investigated
  in the two-nucleon system~\cite{CalleCordon:2009pit},
  where its main advantage is that it allows for
  the {\it renormalization} of the OBE model,
  i.e. it becomes possible to generate observable results
  that are independent of meson form factors and cutoffs.
  The detailed study of these effects lies however beyond the scope of
  the present work: for instance, it will involve the determination of
  four independent coupling constants corresponding to
  the four independent potential components
  of Table \ref{tab:potential-XZ}, or a minimum of four molecular
  candidates to calibrate said couplings.
}

\begin{table}[!h]
\begin{tabular}{|ccc|}
  \hline \hline
  Coupling  & Value & Relevant to meson(s) \\
  \hline
  $g_1$ & 0.60 & $\pi$, $a_1$\\
  $g_{\sigma}$ & 3.4 & $\sigma$ \\
  $g_{V 1}$ & 2.9 & $\rho$, $\omega$ \\
  $\kappa_{V 1}$ & 2.8 & $\rho$, $\omega$ \\
  $\lambda_1$ & 1.8 & $a_1$ \\
  \hline \hline
\end{tabular}
\caption{Couplings of the light-mesons we are considering in this work
  ($\pi$, $\sigma$, $\rho$, $\omega$ and $a_1$) to the charmed mesons.
  For the masses of the light-mesons we will use $m_{\pi} = 138\,{\rm MeV}$,
  $m_{\sigma} = 550\,{\rm MeV}$, $m_{\rho} = 770\,{\rm MeV}$,
  $m_{\omega} = 780\,{\rm MeV}$ and $m_{a_1} = 1230\,{\rm MeV}$.
  For the vector mesons we use the scaling mass $M = 938\,{\rm MeV}$.
  For the charmed mesons we will consider their isospin-averaged masses,
  $m_D = 1867\,{\rm MeV}$ and $m_{D^*} = 2009\,{\rm MeV}$.
}
\label{tab:couplings}
\end{table}

\subsection{Form-factors and Regulators}
\label{subsec:ff}

As mentioned, the previous form of the potential assumes point-like hadrons.
The finite size of the hadrons involved can be taken into account
with different methods, e.g. form factors.
The inclusion of form factors amounts to multiply each vertex involving
a heavy hadron and light meson by a function of the exchanged momentum,
i.e.
\begin{eqnarray}
  \mathcal{A}_R(H \to H M(q)) = f_M({q})\,\mathcal{A}(H \to H M(q)) \, , 
\end{eqnarray}
where $\mathcal{A}$ and $\mathcal{A}_R$ are the point-like and regularized
amplitudes, respectively, and $f(q)$ the form factor.
In terms of the potential, the inclusion of a form factor is equivalent
to the substitution rule
\begin{eqnarray}
  V_M(\vec{q}) \to f_M^2(\vec{q})\,V_M(\vec{q}) \, .
\end{eqnarray}
Here we will use multipolar form-factors, i.e.
\begin{eqnarray}
  f_M(q) = {\left( \frac{\Lambda^2 - m^2}{\Lambda^2 - q^2} \right)}^{n_P} \, ,
  \label{eq:multipolar-FF}
\end{eqnarray}
with $\Lambda$ the form-factor cutoff, $q^2 = -q_0^2 + \vec{q}^2$
the exchanged 4-momentum of the meson $M$, $m$ the mass of said meson
and $n_P$ the multipole momentum.
In general this procedure requires that $\Lambda > m$.

{
  The form-factor cutoff can be different for each of the exchanged mesons,
  which is what happens for instance in the meson theory
  of nuclear forces~\cite{Machleidt:1987hj,Machleidt:1989tm}.
  However, we will also consider the simplification of a single cutoff
  for all exchanged mesons: this choice is popular within OBE descriptions of
  hadronic molecules as it entails less free parameters.
  In contrast with the two-nucleon system, the number of actual data
  for the different two-hadron systems is usually limited to a few
  bound state candidates at most.
  This indeed favors theoretical simplifications such as a single cutoff,
  but it is important to stress that there is no compelling phenomenological
  reason why this should have to be the case.
}
For the inclusion of the axial meson, {we advance that the assumption of
  a single cutoff} entails $n_P \geq 2$ {(otherwise, the cutoff
  will be smaller than the axial meson mass)}.
{
  If we allow for each meson to have its own cutoff,
  there will be no constraints on the polarity of
  the form factor.
  
  It is also worth mentioned that in a first approximation
  we will assume the cutoffs to be identical in the different
  isospin, flavor and heavy/light-quark spin channels.
  However this assumption only holds if the previous symmetries
  are perfectly preserved, which is not the case.
  We will later discuss how the breaking effects of these symmetries
  (in particular flavor and HQSS) might play a role in the coherent
  description of the $X(3872)$, $Z_c(3900)$
  and also the $Z_{cs}(3985)$.
}

Multipolar form factors are local regulators and thus they still generate
a local potential for which the Fourier-transform is analytic.
The expressions can be a bit convoluted though, particularly
for the dipolar and higher momentum form factors.
Here we only consider the S-wave piece of the light-meson exchange potentials
where the contact-range contributions have been removed,
which slightly simplifies the analytic expressions.
First, for each meson we have a Yukawa-like potential of the type
\begin{eqnarray}
  V_M(\vec{r}) &=& \frac{g_Y^2}{4 \pi}
  \left( \mathcal{O}_I \sum_{i} c_{i}  \mathcal{O}^i_S \right)\,
  \frac{e^{-m r}}{r} \nonumber \\
  &=& g_Y^2 \left( \mathcal{O}_I \sum_{i} c_{i}  \mathcal{O}^i_S \right)\,
  m\,W_Y(m r) \, ,
\end{eqnarray}
with $g_Y$ the effective Yukawa coupling, $\mathcal{O}_I$ and $\mathcal{O}_S$
isospin and spin operators, $m$ the mass of the exchanged light-meson and
where the exact potential could involve a sum of different spin operators
with $c_i$ their coefficients.
In the second line we have included the dimensionless function
\begin{eqnarray}
  W_Y(x) = \frac{e^{- x}}{4 \pi x} \, ,
\end{eqnarray}
which is the only thing that changes when a multipolar form factor is included
\begin{eqnarray}
  W_Y(x) \to W_Y(x,\lambda; k_P) \, ,
\end{eqnarray}
where $\lambda = \Lambda / m$, $k_P = 2 n_P$ and with
\begin{eqnarray}
W_Y(x,\lambda, k_P) = \int \frac{d^3 z}{(2\pi)^3}\,
  {\left(\frac{\lambda^2-1}{\lambda^2 + z^2}\right)}^{k_P}\,
  \frac{e^{i \vec{z} \cdot \vec{x}}}{1+z^2}\, .
\end{eqnarray}
The general form of $W_Y$ for integer $k_P \geq 1$ can be found by recursion.
If we redefine $W_Y$ as
\begin{eqnarray}
  W_Y(x,\lambda; k_P) = (\lambda^2 - 1)^{k_P}\,I_Y(x,\lambda; k_P) \, ,
\end{eqnarray}
then $I_Y$ follows the recursive relation
\begin{eqnarray}
  I_Y(x,\lambda; 1) &=& \frac{1}{\lambda^2-1}\,\left[
    W_Y(x) - \lambda W_Y(\lambda x) \right] \, , \\
  I_Y(x,\lambda; k_P > 1) &=& \frac{1}{2 \lambda\, k_P}\,
  \frac{d}{d \lambda}\,\left[ I_Y(x,\lambda; k_P - 1) \right] \, ,
\end{eqnarray}
from which we can find the form of the potential for arbitrary multipolar
form factors.

For the choice of the polarity $n_P$, {if we decide to follow
  the simplifying assumption of a single cutoff for all
  exchanged mesons, we will} have to choose at least a dipolar
form factor ($n_P \geq 2$).
Otherwise the form factor cutoff will be lighter than the axial meson,
rendering it impossible the inclusion of said meson
with a multipolar form factor.
In particular for $n_P = 1$ we indeed have to remove the axial meson to
correctly reproduce the $X(3872)$, in which case the necessary cutoff 
is $\Lambda = 1.00\,{\rm GeV}$ (compatible with
$\Lambda = 1.01^{+0.18}_{-0.10}\,{\rm GeV}$ in the OBE of Ref.~\cite{Liu:2019stu},
which also uses a monopolar form factor).

\subsection{Finite meson width effects}
\label{subsec:meson-width}

Previously we have simply assumed that the light mesons generating the OBE
potential are narrow states.
However, though this assumption is well justified in the case of the pion,
the omega and to a lesser extent the rho, the scalar and axial mesons are
rather broad actually, where in the RPP~\cite{Zyla:2020zbs}
their widths are listed to be:
\begin{eqnarray}
  \Gamma_{\sigma} = 400-800\,{\rm MeV} \quad \mbox{and} \quad
  \Gamma_{a_1} = 250-600\,{\rm MeV} \, . \nonumber \\
\end{eqnarray}
Naturally this raises the question of how these large widths can be taken
into account within the OBE model.

This problem being rather conspicuous, particularly in the case of
the $\sigma$, has been investigated in the past.
The basic idea is to substitute the narrow meson propagator by a propagator
averaged over the actual mass distribution of the meson
\begin{eqnarray}
  \frac{1}{m^2 + {\vec{q}\,}^2} \to
  \int_{m_{\rm th}}^{\infty} \frac{\rho(\mu^2)\,d (\mu^2)}{\mu^2 + {\vec{q}\,}^2} \, ,
  \label{eq:mass-int}
\end{eqnarray}
where $\rho(\mu^2)$ is the spectral distribution of the wide meson and
$m_{\rm th}$ the threshold mass of the particles
into which this meson can decay.
The evaluation of this integral depends though on the form of $\rho(\mu^2)$,
which in turn has led to different approximations for handling wide mesons.
Five decades ago Ref.~\cite{Binstock:1971duy} proposed a really
practical solution for wide mesons decaying into two pions
(i.e. this solution is tailored for the $\sigma$),
which amounts to a two-pole approximation of
the previous integral
\begin{eqnarray}
  \int_{m_{\rm th}}^{\infty} \frac{\rho(\mu^2)\,d (\mu^2)}{\mu^2 + {\vec{q}\,}^2}
  \approx
  \frac{\alpha_1}{m_1^2 + {\vec{q}\,}^2} +
  \frac{\alpha_2}{m_2^2 + {\vec{q}\,}^2} \, ,
\end{eqnarray}
where $\alpha_1$ and $\alpha_2$ are positive numerical coefficients
such that $\alpha_1 + \alpha_2 = 1$ and $m_1$, $m_2$ are
the masses of the two poles, which obey
the relation $m_1 < m$ and $m_2 > m$.
This two-pole approximation has been used for instance in the Nijmegen
high precision potentials~\cite{Stoks:1994wp}.

Later, Ref.~\cite{Flambaum:2007xj} has proposed a more detailed solution,
which begins by considering the exchange of a narrow meson
generating a Yukawa-like potential of the type
\begin{eqnarray}
  V_Y(r) = - \frac{g_Y^2}{4\pi} \frac{e^{-m r}}{r} \, ,
\end{eqnarray}
with $g_Y$ the coupling.
When this meson acquires a width, the evaluation of the integral
over the spectral mass distribution (i.e. Eq.~(\ref{eq:mass-int}))
results in a potential that is the sum of a few contributions,
where we refer to Ref.~\cite{Flambaum:2007xj} for details.
The two most important of these contributions are the following:
close to $m r \sim 1$, the original Yukawa potential is modified
into a potential of the type
\begin{eqnarray}
  V_Y(r) \approx - \frac{g^2}{4\pi}\,
  \frac{e^{-m r}}{r}\,(1 - \frac{\Gamma r}{\pi} -
  \frac{\Gamma}{\pi m}) \, ,
\end{eqnarray}
with $\Gamma$ the width of the meson.
In practical terms this implies that at short distances the potential
for a broad meson is weaker than for a narrow one, where it is
interesting to notice that one can still use
the pole mass of the meson.
The other relevant contribution to the potential of a broad meson comes from
this meson decaying into two lighter mesons of mass $2 M$, which will
generate an additional attractive longer-range contribution to
the potential at distances $2 M r \sim 1$.
In the case of the $\sigma$ meson, we would expect the appearance of a
two-pion exchange contribution, while for the $a_1$ meson
this procedure will give us a $\pi \rho$ exchange
potential.
To summarize, in comparison to a narrow meson, a wide meson exchange potential
will be weaker at short distances and stronger at long distances.

At this point it is worth noticing that though
the analysis of Ref.~\cite{Flambaum:2007xj} has not been explicitly used
for the construction of meson exchange potentials,
it nonetheless explains the features of previous OBE models.
For instance, the relation between two-pion and sigma exchange~\cite{Holzenkamp:1989tq,Reuber:1995vc,Oset:2000gn,CalleCordon:2009ps}
(or $\pi \rho$ and $a_1$ exchange~\cite{Durso:1977ek,Durso:1984um})
is well known and has been
extensively discussed in the past.

Yet, the actual effect of a meson is related both to its coupling and range.
In this regard, it might be possible to simply disregard the complications
coming from the width of the meson in favor of considering the mass and
coupling of said meson as effective parameters.
This idea seems to be compatible with the practice of OBE models, where
we can compare the Bonn-A/B~\cite{Machleidt:1987hj,Machleidt:1989tm}
and CD-Bonn~\cite{Machleidt:2000ge} models for illustrative purposes.
The Bonn-A and -B models are traditional OBE potentials where the exchanged
bosons are effectively treated as narrow: for the case of the $\sigma$ meson
it features a mass and a coupling of $m_{\sigma} = 550\,{\rm MeV}$ and
$g_{\sigma}^2 / 4 \pi \simeq 8.7-8.8$ (which basically coincides with
coupling expected in the L$\sigma$M, $g_{\sigma}^2 / 4 \pi \simeq 8.3$).
In contrast the CD-Bonn model includes two-boson exchange
(e.g. two-pion exchange) and the mass and couplings of the $\sigma$ meson
are $m_{\sigma} = 452\,{\rm MeV}$ and $g_{\sigma}^2 / 4 \pi \simeq 4.25$
instead.
This actually illustrates a few of the ideas of Ref.~\cite{Flambaum:2007xj}
in practice: the strength of the $\sigma$ exchange potential in the CD-Bonn
model is about $50\%$ weaker than in the Bonn-A/B model,
but this is counterbalanced by the presence of
two-pion exchange in CD-Bonn.
At the end the two types of OBE potentials are roughly equivalent at
the observable level, yielding comparable predictions.

From this later observation, in this work we will simply treat the wide
scalar and axial mesons as narrow, with their masses and couplings being
considered as {\it effective} instead of fundamental.
Obviously there is a relation between the effective and physical values,
which is what is discussed in Refs.~\cite{Binstock:1971duy,Flambaum:2007xj},
but no unique solution for taking into account
the finite width effects.
This in turn explains why the parameters of these mesons can take such
different values depending on the approach.
What we will do then is to consider how predictions change with different
values of the masses and couplings of the scalar and axial mesons.

\subsection{Mass gaps and effective meson masses}
\label{subsec:mass-gaps}

When the light-meson exchange potential entails a transition
between two hadrons of different masses, i.e.
there is a vertex of the type
\begin{eqnarray}
  H \to H' M \, ,
\end{eqnarray}
with $H$, $H'$ the initial and final hadrons and $M$ the light-meson,
we will have to modify the effective mass of
the in-flight light meson.
In this case the light-meson exchange potential is not diagonal
and entails a transition between the $H H'$ and $H' H$
configurations
\begin{eqnarray}
  V_M(\vec{q}) = V_M(\vec{q}, HH' \to H'H) \, ,
\end{eqnarray}
for which the light-meson propagator in the exchange potential
will change to
\begin{eqnarray}
  \frac{1}{m^2 + {\vec{q}\,}^2} \to \frac{1}{\mu^2 + {\vec{q}\,}^2} \, ,
\end{eqnarray}
where $\mu$ is the effective mass of the light-meson, i.e.
\begin{eqnarray}
  \mu^2 = m^2 - \Delta^2 \, ,
\end{eqnarray}
with $\Delta = m(H') - m(H)$.

If we are dealing with a charmed meson-antimeson system, this correction will
only have to be taken into account for the $D^* \bar{D}$ and $D \bar{D}^*$
systems, i.e. for the $X(3872)$ and $Z_c(3900)$.
Only the spin-spin part of the potential will be affected, as the central part
cannot generate a transition between the $D$ and $D^*$ charmed mesons.
For the pion and axial meson exchange potentials, the correction is trivial
\begin{eqnarray}
  V_{\pi}(\vec{r}) &=& \zeta\,\frac{g_1^2 \mu_{\pi}^2}{6 f_{\pi}^2} \,
  \vec{\tau}_1 \cdot \vec{\tau}_2 \,
  \vec{\sigma}_{L1} \cdot \vec{\sigma}_{L2} \,
  \frac{e^{-\mu_{\pi} r}}{4 \pi r} \, , \label{eq:V-pi-delta} \\ 
V_{a_1}(\vec{r}) &=& -\zeta\, \lambda_1^2 \frac{g_1^2 m_{a1}^2}{3 f_{\pi}^2} \,
  \vec{\tau}_1 \cdot \vec{\tau}_2 \,
  \vec{\sigma}_{L1} \cdot \vec{\sigma}_{L2} \, \frac{e^{-\mu_{a1} r}}{4 \pi r} \, ,
  \label{eq:V-a1-delta}
\end{eqnarray}
with $\mu_{\pi}$ and $\mu_{a1}$ the effective pion and axial meson masses, where
we notice that for axial meson exchange the $m_{a1} \to \mu_{a1}$ substitution
is limited to the long-range decay exponent (but not the $m_{a1}^2$ factor
involved in the strength of the potential).
For the vector meson exchange potential the correction only affects
its spin-spin piece
\begin{eqnarray}
  V_V(\vec{r}) &=& g_{V1}^2 \frac{e^{-m_V r}}{4 \pi r} \nonumber \\ &+&
  f_{V1}^2\frac{\mu_V^2}{6 M^2}\,\vec{\sigma}_{L1} \cdot \vec{\sigma}_{L2} \, \frac{e^{-\mu_V r}}{4 \pi r} \, .
\end{eqnarray}
Finally for the sigma meson no modification is required.

If we combine this modification with a multipolar form factor,
from direct inspection of Eq.~(\ref{eq:multipolar-FF})
we find that besides modifying the effective mass of the light-meson,
we also have to modify its cutoff
\begin{eqnarray}
  m \to \mu \quad \mbox{and} \quad \Lambda \to \sqrt{\Lambda^2 - \Delta^2} \, .
\end{eqnarray}
Taking into account that $m(D^*) - m(D) \sim 140\,{\rm MeV}$,
the only light-meson that is substantially affected by
this change is the pion, for which its spin-spin contribution
essentially vanishes for the $X(3872)$ and $Z_c(3900)$ molecules.

\subsection{Bound state equation}

For obtaining predictions with the S-wave OBE potential of
Eq.~(\ref{eq:OBE-potential}), we plug it
into the reduced Schr\"odinger equation
\begin{eqnarray}
  -u''(r) + 2\mu_{HH}\,V_{\rm OBE}(r) u(r) = - \gamma^2 \, u(r) \, , 
\end{eqnarray}
where $u(r)$ is the reduced wave function, $\mu_{HH}$ the reduced mass of
the particular charmed meson-antimeson system under consideration and
$\gamma$ the wave number of the bound state, which is related to
the two-body binding energy ($B_2$) by $B_2 = \frac{\gamma^2}{2\mu_{HH}}$.
For the $X(3872)$, we will consider it to be a $D^*\bar{D}$ system
bound by $4\,{\rm MeV}$ in the isospin symmetric limit (i.e. we will be using
the isospin averaged $D$ and $D^*$ masses).
For the $Z_c(3900)$ we will be interested in determining the cutoff for which
it becomes a bound state at threshold, that is, the $\gamma = 0$ limit,
as we will explore in the following lines.

\subsection{The $X(3872)$ and $Z_c(3900)$ cutoffs}

{
Now we will explore whether the $X(3872)$ and $Z_c(3900)$ can be explained
with the same set of parameters in a simplified OBE model
with a single cutoff for all exchanged mesons.
This is not necessarily the most realistic assumption --- the OBE model
in the two-nucleon sector has different cutoffs for each of
the exchanged meson~\cite{Machleidt:1987hj,Machleidt:1989tm} ---
but it allows for a simpler analysis.
Here what we are interested in is whether it is plausible to explain these
two states with compatible parameters in the molecular picture,
which seems to be the case.
}

{
As previously discussed, the single cutoff OBE model requires a polarity
of $n_P \geq 2$ for the form factor.
}
With a dipolar form factor ($n_P = 2$) we can reproduce the mass of the $X$ with
\begin{eqnarray}
  \Lambda(X) = 1.37^{+0.08}_{-0.09}\,(1.27-1.41)\,{\rm GeV} \, .
  \label{eq:Lambda-X}
\end{eqnarray}
The central value corresponds to $m_{\sigma} = 550\,{\rm MeV}$, which is
the $\sigma$ mass used in the original OBE model for nuclear
forces~\cite{Machleidt:1987hj,Machleidt:1989tm},
the error comes from the uncertainty in the scalar coupling
$g_{\sigma} = 3.4 \pm 0.1$, while the spread represents the mass range
$m_{\sigma} = 450-600\,{\rm MeV}$ (with $g_{\sigma} = 3.4$),
which covers most of the plausible choices for its mass.
For $m_{\sigma} = 450\,{\rm MeV}$ the dipolar cutoff is merely a bit above
the axial meson mass, $m_{a1} = 1.23\,{\rm GeV}$.
In contrast the cutoff for which the $Z_c$ binds at threshold is
\begin{eqnarray}
  \Lambda({Z}_c) = 1.82^{+0.62}_{-0.43}\,(1.37-1.99)\,{\rm GeV} \, ,
\end{eqnarray}
for which we get the ratio
\begin{eqnarray}
  \frac{\Lambda(Z_c)}{\Lambda(X)} = 1.33^{+0.35}_{-0.24}\,(1.08-1.41) \, .
\end{eqnarray}
Now if we assume that the cutoffs for the $X$ and the $Z_c$ are the same,
modulo HQSS violations {(the size of which is about $\Lambda_{\rm QCD} / m_Q$,
  i.e. of the order of $15\%$ for $m_Q = m_c$, where we have taken $\Lambda_{\rm QCD} \sim 200\,{\rm MeV}$)}, this ratio should be one
within the aforementioned HQSS uncertainty
\begin{eqnarray}
  \frac{\Lambda_Z}{\Lambda_X} = 1 \pm 0.15 = (0.85-1.15) \, ,
\end{eqnarray}
which means that the existence of the $X$ is compatible within one standard
deviation with a $Z_c$ binding at threshold for the lower
$\sigma$ meson mass range.
This lower $\sigma$ mass range basically gives cutoffs that are barely
larger than the axial meson mass, which indicates that we should consider
larger dipolar momenta for a better comparison.
This is done in Table~\ref{tab:cutoff-XY}, where we extend the comparison to
the $n_P = 3, 4$ (i.e. tripolar and quadrupolar) cases.
Yet we consistently end up about two standard deviations away.
However this discrepancy is not troubling: a molecular $Z_c$ is expected to be
a virtual state or a resonance near threshold,
which means that the amount attraction in the $1^{+}(1^{+-})$ $D^*\bar{D}$ system
is not enough to bind the charmed meson and antimeson at threshold.
That is, the ratio should be larger than one (but still of $\mathcal{O}(1)$),
though it is difficult to estimate how much larger.
Thus the previous ratio and the ones in Table~\ref{tab:cutoff-XY} are probably
compatible with a molecular $Z_c$.

\begin{figure*}[ttt]
\begin{center}
  \includegraphics[width=8.9cm]{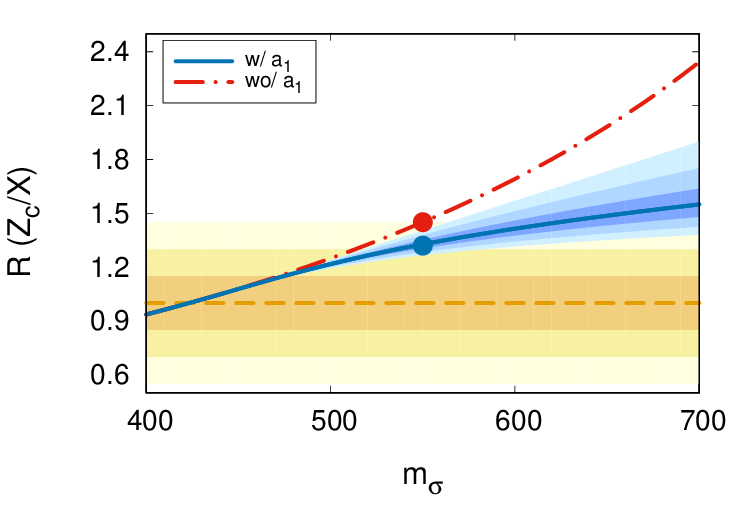}
  \includegraphics[width=8.9cm]{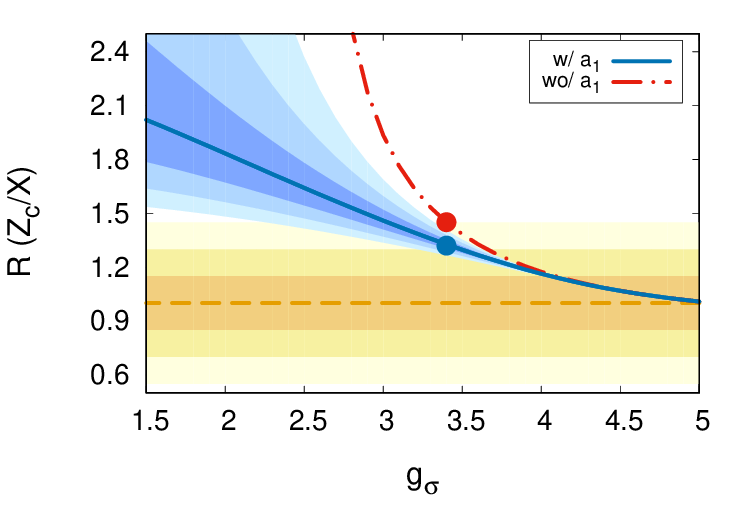}
\end{center}
\caption{
  Cutoff ratios $R(Z_c/X)$ as a function of the mass and the coupling of
  the scalar meson for the OBE model with (solid lines) and
  without (dashed-dotted lines)   axial meson exchange.
  $R(Z_c/X)$ is defined as the ratio of the cutoff for which a molecular $Z_c$
  will be a bound state at threshold over the cutoff for which the mass of
  the $X(3872)$ is reproduced as a $I^G(J^{PC}) = 0^{+}(1^{++})$ $D^* \bar{D}$
  bound state.
  If the $Z_c$ were to be a bound state at threshold, the ratio would be
  expected to be $R(Z_c/X) = 1.0 \pm 0.15$, which we show in the figure
  as a dashed line and a series of bands representing one, two and three
  standard deviations (shown in increasingly light colors).
  When we vary the scalar meson mass (coupling), the scalar coupling (mass)
  is taken to be $g_{\sigma} = 3.4$ ($m_{\sigma} = 550\,{\rm MeV}$).
  The predictions for the preferred values of the scalar mass and
  couplings ($m_{\sigma} = 550\,{\rm MeV}$ and $g_{\sigma} = 3.4$)
  are highlighted as round dots.
  For the axial meson, we use $m_{a_1} = 1230\,{\rm MeV}$ and
  $\lambda_1 = 1.8 \pm 0.3$, where the uncertainty in the axial coupling
  is shown as a series of bands around the central predictions
  for the axial-full theory (where we show again the one, two and
  three standard deviations bands in increasingly light colors).
}
\label{fig:R-scalar}
\end{figure*}

However the most interesting comparison is against the OBE model
without the axial meson, which will reveal the conditions
under which the axial meson might be relevant.
If we remove the axial meson, the cutoffs we get are
\begin{eqnarray}
  \Lambda_X^{\slashed{a}_1} &=& 1.37^{+0.09}_{-0.09}\,(1.27-1.41)\,{\rm GeV} \, , \\
  \Lambda_Z^{\slashed{a}_1} &=& 1.99^{+\infty}_{-0.60}\,(1.37-2.38)\,{\rm GeV} \, ,
\end{eqnarray}
where $Z_c$ does not bind for the lower values of $g_{\sigma}$ (we discuss this
in a moment), hence the $^{+\infty}$ error, yielding the ratio
\begin{eqnarray}
  \frac{\Lambda_Z^{\slashed{a}_1}}{\Lambda^{\slashed{a}_1}_X} =
  1.45^{+\infty}_{-0.36}\,(1.08 -1.69) \, ,
\end{eqnarray}
which results in larger relative cutoffs as the $\sigma$ gets heavier,
increasing the discrepancy with HQSS to the three standard deviation level
(if we assume a molecular $Z_c$ at threshold, which is probably
too restrictive). The ratios also grow larger for higher polarity $n_P$,
see Table \ref{tab:cutoff-XY}.
Again lighter $\sigma$ masses result in cutoffs that are not completely
satisfactory if we want to take the axial mesons into account.
Finally in the left panel of Fig.~\ref{fig:R-scalar} we show the dependence of
the cutoff ratio with the mass of the $\sigma$ for a dipolar form factor,
where we can see again that the impact of the axial meson increases
with the mass of the scalar meson.

At first sight the comparison between the axial-full and axial-less OBE models
indicates a modest contribution from the axial mesons.
But the observation that the cutoff ratios increase with larger scalar meson
masses, left panel of Fig.~\ref{fig:R-scalar}, and with it
the compatibility of the molecular description of the $X$ and
the $Z_c$ decreases, indicates that the previous conclusion
depends on the strength of scalar meson exchange and
the parameters used to describe the later.
We actually do not know the coupling of the $\sigma$ to the charmed baryons
very precisely, but with considerable errors: the L$\sigma$M and
the quark model suggest $g_{\sigma} = 3.4 \pm 1.0$, where
this uncertainty turns out to be important.
If the attraction provided by the $\sigma$ falls short of binding, the axial
meson will be the difference between the charmed meson-antimeson interaction
being weak or strong.
Indeed, if there is no axial meson, the condition for the isovector
$D^*\bar{D}$ system to bind is
\begin{eqnarray}
  g_{\sigma} \geq 2.45\,(2.22 - 2.56)\, ,
\end{eqnarray}
which is within the expected uncertainties for the scalar coupling.
That is, $\sigma$ exchange is by itself no guarantee that the $Z_c(3900)$
can be explained in terms of the charmed meson-antimeson
interaction alone.
In the right panel of Fig.~(\ref{fig:R-scalar}) we visualize
the dependence of the $\Lambda(Z_c) / \Lambda(X)$ ratio as a
function of $g_{\sigma}$,
which further supports the previous interpretation of
axial meson exchange as the factor guaranteeing
the required molecular interaction necessary
for the $Z_c$.
Finally for a $\sigma$-less theory with axial mesons the $Z_c$ will still
bind for large cutoffs, with concrete calculations yielding
\begin{eqnarray}
  \Lambda^{\slashed{\sigma}}(X) = 1.55\,{\rm GeV} \quad \mbox{and} \quad
  \Lambda^{\slashed{\sigma}}(Z_c) = 3.62\,{\rm GeV} \, ,
\end{eqnarray}
for which the ratio is $2.33$. In this later scenario the uncertainty of
the factor involved in the mixing of the pion and axial meson current
($\lambda_1 = 1.8 \pm 0.3 = 1.5-2.1$) might be relevant, as this error
induces the ratio to move within the $2.01-2.99$ range. 

\begin{figure*}[ttt]
\begin{center}
  \includegraphics[width=8.9cm]{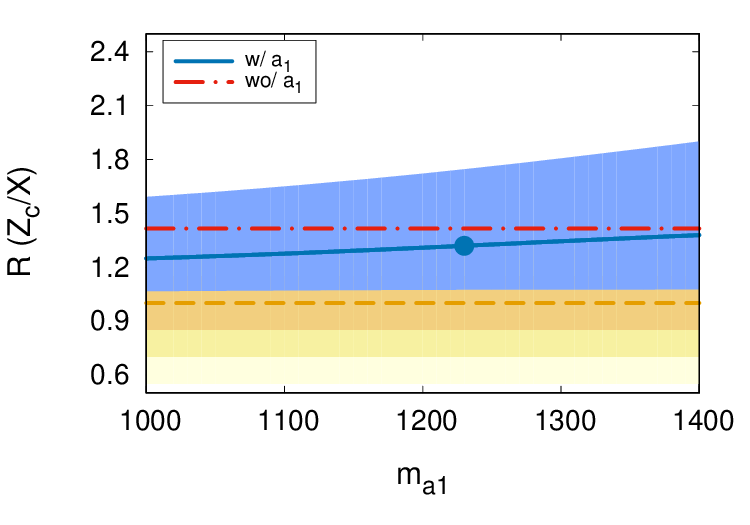}
  \includegraphics[width=8.9cm]{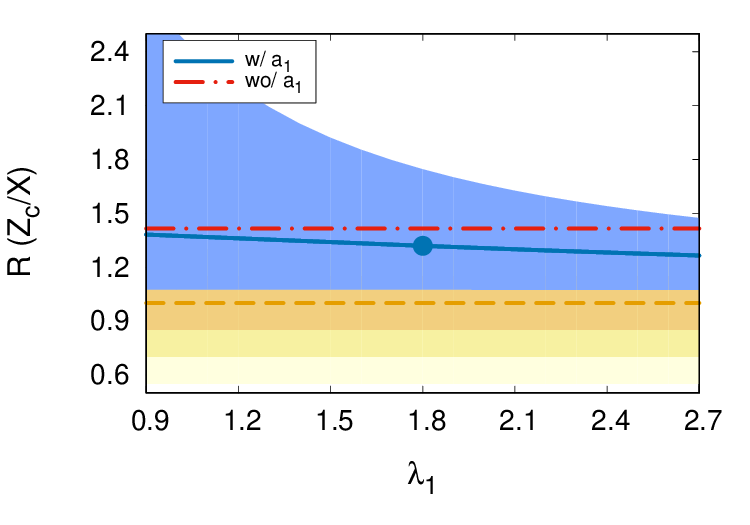}
\end{center}
\caption{
  Cutoff ratios $R(Z_c/X)$ as a function of the mass (left panel) and
  coupling (right panel) of the axial meson for the OBE model.
  $R(Z_c/X)$ is defined as in Fig.~\ref{fig:R-scalar}, with its expected value
  being $R(Z_c/X) = 1.0 \pm 0.15$, which we show as a dashed line and
  a series of bands representing one, two and three
  standard deviations (shown in increasingly light colors).
  The solid and dashed-dotted lines represent the ratios for the OBE with
  and without axial meson exchange.
  The mass and coupling of the scalar meson are taken to be
  $m_{\sigma} = 550\,{\rm MeV}$ and $g_{\sigma} = 3.4 \pm 1.0$,
  where in the case of the coupling we have added a $30\%$
  relative uncertainty, which is shown as the error band
  around the solid line.
  In contrast with Fig.~\ref{fig:R-scalar}, where we showed up to three standard
  deviations of the uncertainty in $\lambda_1$, for $g_{\sigma}$ the errors are
  considerably larger and we show only one standard deviation instead.
  For the axial-less OBE model the ratio is constant, as there is no dependence
  with respect to the axial meson parameters, and calculated to be
  $R^{\slashed a_1} = 1.42^{+\infty}_{-0.34}$, where the error comes
  from the uncertainty of $g_{\sigma}$ (not shown in the plots
  owing to its large spread).
  The $m_{a_1} = 1230\,{\rm MeV}$ and $\lambda_1 = 1.8$ ratios for the axial-full
  theory are highlighted as a round dot. When we vary the axial mass 
  (coupling), we set the axial coupling (mass) to its expected central
  value, i.e. $\lambda_1 = 1.8$ ($m_{a_1} = 1230\,{\rm MeV}$).
}
\label{fig:R-axial}
\end{figure*}

{
The previous analysis of the scalar mass and coupling dependence is motivated
in part by the width of the $\sigma$, which implies that its parameters
as applied to the OBE model are in a sense {\it effective}
and do not necessarily coincide with its physical parameters,
see the discussion in Sect.~\ref{subsec:meson-width}.
This factor is also present for the axial meson, though in a lesser extent
owing to a smaller width to mass ratio.
Nonetheless it is relevant to study how the cutoff ratio depends on
$m_{a_1}$ and $\lambda_1$, which we do in Fig.~\ref{fig:R-axial}.
What we find is that the uncertainties in the $R(Z_c/X)$ ratio show a weaker
dependence with the axial meson mass and coupling
with respect to the scalar meson.
This result is in line with the previous observations that the importance of
axial meson exchange is indeed subordinated to the scalar meson, which
having a larger range naturally exerts a larger influence of
the spectrum.
}

{
  Finally, we remind that here we have worked under the assumption that
  the $\Lambda_X$ and $\Lambda_Z$ cutoffs should be identical.
  This would be true if the form-factor cutoffs of the OBE model are determined
  by the particles for which the potential is calculated (i.e. the charmed
  mesons), instead of by the light mesons being exchanged.
  However, this is not necessarily the case: in the OBE model of nuclear
  forces~\cite{Machleidt:1987hj} the cutoffs of the light-mesons are of
  the same order of magnitude, but not identical.
  This leads us to a second interpretation: the use of a unique cutoff could
  then be regarded as an average of the particular cutoffs of each of
  the light mesons.
  If this is the case, the $\Lambda_X$ and $\Lambda_Z$ cutoffs could differ
  from each other much more than naively expected, where a very important
  reason for this is that in the isovector molecules there is no vector
  meson exchange, giving a much larger weight for the axial meson
  cutoff in the $Z_c$-like molecules.
  In turn this will allows us to make independent predictions of
  the isoscalar and isovector charmed meson-antimeson molecules.
  We will explore this possibility in the following lines.
}

\begin{table*}[!ttt]
\begin{tabular}{|c|cc|cc|cc|}
\hline\hline
Polarity ($n_P$) &
$\Lambda(X)$ & $\Lambda^{\slashed{a}_1}(X)$ &
$\Lambda(Z_c)$ & $\Lambda^{\slashed{a}_1}(Z_c)$ &
$R(Z_c/X)$ & $R^{\slashed{a}_1}(Z_c/X)$ \\
\hline
2 & $1.37^{+0.08}_{-0.09}$ (1.27-1.41) & $1.37^{+0.09}_{-0.09}$ (1.27-1.41) & $1.82^{+0.62}_{-0.43}$ (1.37-1.99) & $1.99^{+\infty}_{-0.60}$ (1.37-2.38)
& $1.33^{+0.35}_{-0.24}$ (1.08-1.41) & $1.45^{+\infty}_{-0.36}$ (1.08-1.69) \\
3 & $1.65^{+0.10}_{-0.11}$ (1.53-1.69) & $1.65_{-0.11}^{+0.11}$ (1.53-1.70) & $2.19^{+0.76}_{-0.51}$ (1.67-2.40) & $2.44_{-0.75}^{+\infty}$ (1.68-2.92)
& $1.33^{+0.36}_{-0.24}$ (1.09-1.42) & $1.48_{-0.38}^{+\infty}$ (1.10-1.72) \\
4 & $1.95^{+0.11}_{-0.13}$ (1.82-2.00) & $1.97^{+0.13}_{-0.14}$ (1.83-2.03) & $2.72^{+0.96}_{-0.65}$ (2.10-2.96) & $3.26^{+\infty}_{-1.16}$ (2.16-4.02)
& $1.39^{+0.39}_{-0.26}$ (1.15-1.48) & $1.65^{+\infty}_{-0.51}$ (1.18-1.98) \\
  \hline\hline
\end{tabular}
\caption{Cutoffs required to reproduce the $X(3872)$ and to bind a molecular
  $Z_c$ at threshold in a OBE model with and without axial mesons
  for different masses of the scalar meson.
  We use a multipolar form factor with polarity $n_P =2,3,4$ at each vertex,
  Eq.~(\ref{eq:multipolar-FF}).
  $\Lambda(X/Z_c)$ shows the $X(3872)$ and $Z_c$ cutoffs for the OBE model
  including the axial meson, while for the axial-less case we add
  the superscript ${}^{{\slashed a}_1}$.
  Finally we show the ratio between the $X(3872)$ and $Z_c$ cutoffs.
  The central value represents $m_{\sigma} = 550\,{\rm MeV}$, the error in
  the central values arise from the uncertainty in $g_{\sigma} = 3.4 \pm 1.0$ and
  the intervals (in parentheses) correspond to $m_{\sigma} = 450-600\,{\rm MeV}$.
}
\label{tab:cutoff-XY}
\end{table*}

\subsection{The HQSS partners of the $X(3872)$ and $Z_c(3900)$}

Now we want to explore what type of molecular spectrum is derived
when the exchange of the axial meson is included.
For this we will consider a two cutoff model where the isoscalar and
isovector configurations are independent, i.e. we will use
the $\Lambda_X$ and $\Lambda_Z$ cutoffs calculated
in the previous section.

Ideally, the description of the two isospin channels would involve
using a different cutoff for each of the exchanged meson, as done
for instance in the meson theory of nuclear forces.
However, instead of calibrating all these parameters --- we have two molecular
candidates only --- here we will consider the interpretation in which
the cutoff works as an effective parameter representing
the different mix of mesons contributing
in each channel.
In particular the isovector $Z_c(3900)$ and $Z_c(4020)$ resonances
do not exchange vector mesons, from which it is sensible to expect
a different ``effective'' or ``average'' cutoff
from the one in the $X(3872)$ case.
Provided these two cutoffs are not too far away, which seems to be the case
for most choices of parameters in the axial-full theory, this simplified
description should be able to generate the qualitative features of
the charmed meson-antimeson molecular spectrum.

The spectrum we obtain is summarized in Table \ref{tab:molecules}
for the axial-less and axial-full OBE models.
While for the isoscalar configurations --- the partners of the $X(3872)$ ---
there is no practical difference between including or excluding
the $a_1(1260)$, for the partners of the $Z_c(3900)$
there are important differences.
We summarize our results as follows:
\begin{itemize}
\item[(i)] In the isoscalar sector (the partners of the $X(3872)$)
  there is no significant difference between including or excluding axial
  meson exchange. The spectrum is similar to the one predicted
  in Ref.~\cite{Liu:2019stu} (an axial-less OBE model),
  though here we observe two additional states that are
  worth mentioning:
  \begin{itemize}
  \item[(i.a)] A near threshold $0^{++}$ $D\bar{D}$ virtual state,
  which could correspond to the previously predicted $X(3700)$
  states~\cite{Gamermann:2006nm,Nieves:2012tt}
  or the recent $0^{++}$ bound state predicted
  in the lattice~\cite{Prelovsek:2020eiw}.
  We note that this latter work includes the $D\bar{D}$-$D_s \bar{D}_s$
  coupled channel dynamics, which will provide additional
  attraction not present here.

\item[(i.b)] A negative C-parity (virtual state) partner of the $X(3872)$,
  which might be related to the observation of a state with these quantum
  numbers by COMPASS~\cite{COMPASS:2017wql}
  with $M = 3860.0 \pm 10.4\,{\rm MeV}$.
  We notice that by taking $g_{\sigma} = 2.4$ we will predict
  the central value of the COMPASS observation.
  This might also indicate that smaller values of the scalar meson coupling
  are to be preferred.
  
  \end{itemize}
\item[(ii)] In the isovector sector (the partners of the $Z_c(3900)$)
  there are interesting difference between the axial-less and axial-full
  models:
  \begin{itemize}
  \item[(ii.a)] In the axial-less model, all the isovector molecules
    have the same binding energy (except for perturbative corrections
    from pion exchanges), i.e. they are all located close to threshold.
    However this conclusion depends on $g_{\sigma} > 2.45$, which is within
    the error that we expect for the scalar coupling
    ($g_{\sigma} = 3.4 \pm 1.0$).
    If this condition is met, there should be a total of six isovector molecules
    or, equivalently, four unobserved partners of the $Z_c(3900)$
    and $Z_c(4020)$.
    On the contrary, if $g_{\sigma}$ is not strong enough
    isovector molecular states will not exist.

\item[(ii.b)] In the axial-full model, $a_1$-exchange generates a spin dependent
  interaction that breaks this degeneracy. The most attractive configuration
  is the $I^G(J^{PC}) = 1^{-}(0^{++})$ $D^*\bar{D}^*$ molecule, followed by
  the $Z_c(3900)$ and $Z_c(4020)$.
  However the size of this effect depends on $g_{\sigma}$:
  for the central value derived from the L$\sigma$M, the hyperfine splitting
  will be of the order of merely $1\,{\rm MeV}$, and the spectrum
  will be barely distinguishable from the axial-less model.
  But if $g_{\sigma}$ is weak, the hyperfine splitting will become sizable.
  For instance, if $g_{\sigma} = 2.4$ the $0^{++}$ $D^*\bar{D}$
  molecule will be more bound than the $Z_c(4020)$
  by about $10\,{\rm MeV}$.
  \end{itemize}
\end{itemize}
In general, it is the isovector molecular spectrum which could provide more
information about scalar and axial meson exchange:
the eventual observation of the HQSS partners of the $Z_c(3900)$ and
$Z_c(4020)$ could determine which of the scenarios discussed here is
the one chosen by nature: a degenerate spectrum would be compatible
with a strong scalar meson exchange (or maybe with vector charmonia
exchange~\cite{Dong:2021juy}),
while the discovery of a $0^{++}$ partner of the $Z_c(4020)$ will signal
that axial meson exchange is a relevant part of the description of
isovector molecules.
However, if a $2^{++}$ partner happens to be discovered and it can be shown
that requires more attraction than the $Z_c(4020)$, this will be difficult
to accommodate in the previous molecular explanations, independently
on whether it is grounded on scalar, axial or charmonium exchange.

\begin{table*}[!ttt]
\begin{tabular}{|c|cc|cc|cc|}
\hline \hline
System ($X$-like) & $I^G$ & $J^{PC}$ &
$B^{\slashed{a}_1}$ / $E_V^{\slashed{a}_1}$ & $M^{\slashed{a}_1}$ &
$B^{{a}_1}$ / $E_V^{a_1}$ & $M^{{a_1}}$ \\
\hline
$D\bar{D}$ & $0^{+}$ & $0^{++}$ &
$-0.0^{+0.1}_{-0.7}$ & $3734.4^{+0.0}_{-0.7}$ &
$-0.0^{+0.1}_{-0.8}$ & $3734.4^{+0.0}_{-0.8}$  \\
\hline
  $D^*\bar{D}$ & $0^{+}$ & $1^{++}$ &
  Input & Input &
  Input & Input \\
  $D^*\bar{D}$ & $0^{-}$ & $1^{+-}$ &
  $-4.0^{+3.6}_{-10.3}$ & $3871.8^{+3.6}_{-10.3}$ &
  $-4.2^{+3.9}_{-11.2}$ & $3871.6^{+3.9}_{-11.2}$ \\
  \hline
  $D^*\bar{D}^*$ & $0^{+}$ & $0^{++}$ &
  ${-}^{-}_{-0.3}$ & $-$ &
  ${-}^{-}_{-0.3}$ & $-$ \\
  $D^*\bar{D}^*$ & $0^{-}$ & $1^{+-}$ &
  $-1.0^{+1.0}_{-1.8}$ & $4016.2^{+1.0}_{-1.8}$ &
  $-1.0^{+1.0}_{-1.9}$ & $4016.2^{+1.0}_{-1.9}$ \\
  $D^*\bar{D}^*$ & $0^{+}$ & $2^{++}$ &
  $+3.5^{+0.0}_{-0.1}$ & $4013.7^{+0.1}_{-0.0}$ &
  $+3.5^{+0.0}_{-0.1}$ & $4013.7^{+0.1}_{-0.0}$ \\
  \hline \hline
System ($Z_c$-like) & $I^G$ & $J^{PC}$ &
$B^{\slashed{a}_1}$ / $E_V^{\slashed{a}_1}$ & $M^{\slashed{a}_1}$ &
$B^{{a}_1}$ / $E_V^{a_1}$ & $M^{{a_1}}$ \\
$D\bar{D}$ & $1^{-}$ & $0^{++}$ &
$-0.1^{+0.0}_{-\infty}$ & $3734.3^{0.0}_{-\infty}$ &
$-0.6^{+6.9}_{-13.4}$ & $3733.8^{+0.6}_{-13.4}$ \\
\hline
  $D^*\bar{D}$ & $1^{-}$ & $1^{++}$ &
  $-0.0^{+0.0}_{-\infty}$ & $3875.8^{+0.0}_{-\infty}$ &
  $-1.3_{-16.1}^{+6.0}$ & $3874.5^{+1.3}_{-16.1}$ \\
  $D^*\bar{D}$ & $1^{+}$ & $1^{+-}$ &
  Input & Input &
  Input & Input \\
  \hline
  $D^*\bar{D}^*$ & $1^{-}$ & $0^{++}$ &
  ${+0.0}^{+0.0}_{-\infty}$ & $4017.2^{+0.0}_{-\infty}$ &
  ${+0.3}^{+10.2}_{-0.3}$ & $4016.9^{+0.3}_{-10.2}$ \\
  $D^*\bar{D}^*$ & $1^{+}$ & $1^{+-}$ &
  $+0.0^{+0.0}_{-\infty}$ & $4017.2^{+0.0}_{-\infty}$ &
  $+0.0^{+0.0}_{-0.0}$ & $4017.2^{+0.0}_{-0.0}$ \\
  $D^*\bar{D}^*$ & $1^{-}$ & $2^{++}$ &
  $+0.2^{+0.0}_{-\infty}$ & $4017.0^{+0.2}_{-\infty}$ &
  $-0.5^{+0.7}_{-4.7}$ & $4016.7^{+0.5}_{-4.7}$ \\
  \hline \hline 
\end{tabular}
\caption{
  HQSS partners of the $X(3872)$ and $Z_c(3900)$ states
  in the molecular picture in the axial-less and
  axial-full theories.
  The spectrum is calculated with the OBE model proposed here and a dipolar
  form factor ($n_P = 2$), where the cutoff is determined independently
  in the isoscalar and isovector channels from the condition of
  reproducing the $X(3872)$ (as an isoscalar $J^{PC} = 1^{++}$
  $D^*\bar{D}$ bound state $4\,{\rm MeV}$ below threshold)
  and the $Z_c(3900)$ (as an isovector $J^{PC} = 1^{+-}$ $D^*\bar{D}$
  bound state at threshold).
  With these conditions we obtain the cutoffs
  $\Lambda_X^{\slashed{a}_1} = 1.37^{+0.09}_{-0.09}\,{\rm GeV}$
  and $\Lambda_Z^{\slashed{a}_1} = 1.99^{+\infty}_{-0.60}\,{\rm GeV}$
  in the axial-less model and
  $\Lambda_X^{{a}_1} = 1.37^{+0.08}_{-0.09}\,{\rm GeV}$
  and $\Lambda_Z^{{a}_1} = 1.82^{+0.62}_{-0.43}\,{\rm GeV}$
  in the axial-full model, where the errors correspond to the uncertainty
  in the $\sigma$ meson coupling, $g_{\sigma} = 3.4 \pm 1.0$ (which is also
  propagated into the binding / virtual state energies).
  Owing to the absence of coupled channels, we only obtain bound or virtual
  states: we use the convention of a positive number for the binding energy
  of a bound state and a negative number for the energy of a virtual state,
  while a dash (``-'') indicates the absence of a pole close to threshold.
}
\label{tab:molecules}
\end{table*}

\section{Description of the new $Z_{cs}(3985)$}
\label{sec:Zcs}

Finally we turn our attention to the $Z_{cs}(3985)$,
recently discovered by BESIII~\cite{Ablikim:2020hsk}.
The existence of this resonance can be readily deduced from the $Z_c(3900)$ and
$Z_c(4020)$ and SU(3)-flavor symmetry, as the latter dictates that
the $D^* \bar{D}$ interaction in the $I=1$ isovector channel
is identical to the one in the $D_s^* \bar{D}$
system~\cite{HidalgoDuque:2012pq,Yang:2020nrt}.
However the realization of SU(3)-flavor symmetry in the OBE model
is not automatic and depends on two conditions
\begin{enumerate}
\item[(i)] the pseudo Nambu-Goldstone meson current mixing
  with the octet part of the axial mesons,
\item[(ii)] the scalar meson coupling with similar strengths to the $q=u,d,s$
  light-quarks.
\end{enumerate}

The first condition is required in order for the axial meson exchange to be
non-trivial in the isovector molecules: if the axial mesons form a clear
nonet with almost ideal decoupling of strange and non-strange components,
as happens with the vector mesons, then axial meson exchange
will cancel out in both the $Z_c(3900)$ and $Z_{cs}(3985)$.
We warn though that the status and nature of the axial mesons --- $a_1(1260)$,
$f_1(1285)$, $f_1(1420)$, $K_1(1270)$ --- is not clear: they might be
composite~\cite{Roca:2005nm}, the $f_1(1420)$ might not
exist~\cite{Aceti:2016yeb,Debastiani:2016xgg}, etc.
Here we will not discuss these issues, but simply point out the conditions
under which they will help explain the $Z_c(3900)$ and $Z_{cs}(3985)$.

The second condition is required for SU(3)-flavor symmetry to be respected
between the $Z_c(3900)$ and $Z_{cs}(3985)$: if the scalar $\sigma$ meson
does not coupled with the strange quarks, then a sizable part of the attraction
in the $Z_c(3900)$ system will simply not be present in the $Z_{cs}(3985)$.
As happened with the axial mesons, the nature of the scalar mesons is not
clear either: they might be $q\bar{q}$ or tetraquark or
a superposition of both, the mixing angle between
the singlet and octet components is not known
or it might violate the OZI (Okubo-Zweig-Iizuka) rule.
We will note that the binding of the $Z_{cs}(3985)$ as a hadronic molecule
requires a $\sigma$ that couples with similar strength to the non-strange
and strange light-quarks, which is not implausible.

\subsection{Flavor structure of the potential}

The $D^{(*)}$ and $D_s^{(*)}$ charmed mesons belong to the $\bar{3}$ SU(3)-flavor
representation.
From this the flavor structure of the $D_a^{(*)} \bar{D}_a^{(*)}$ potential,
where $a = 1,2,3$ refers to the $D^0$, $D^+$ and $D_s^+$,
is expected to be $3 \otimes \bar{3} = 1 \oplus 8$,
i.e. the sum of a singlet and octet contributions:
\begin{eqnarray}
  V(D_a^{(*)} \bar{D}_a^{(*)}) = \lambda^{(S)} V^{(S)} + \lambda^{(O)} V^{(O)} \, , 
\end{eqnarray}
with the superscript $^{(S)}$ and $^{(O)}$ referring to the singlet and octet
and where the specific decomposition is shown in Table \ref{tab:su3}.
Of course the flavor structure compounds with the HQSS structure , that is,
the singlet and octet potential can be further decomposed into a central
and a spin-spin part
\begin{eqnarray}
  V^{(S)} &=&
  V_a^{(S)} + V_b^{(S)} \, \vec{\sigma}_{L1} \cdot \vec{\sigma}_{L2} \, ,   \label{eq:V-singlet}
  \\
  V^{(O)} &=&
  V_a^{(O)} + V_b^{(O)} \, \vec{\sigma}_{L1} \cdot \vec{\sigma}_{L2} \, .
  \label{eq:V-octet}
\end{eqnarray}
Finally, the relation between the singlet and octet components and the isospin
components we previously defined for the $X(3872)$ and the $Z_c(3900)$ is
\begin{eqnarray}
  V^{(S)} &=& \frac{3}{2} V^{(I=0)} - \frac{1}{2} V^{(I=1)} \, , \\
  V^{(O)} &=& V^{(I=1)} \, .
\end{eqnarray}

From the flavor decomposition of the potential (Table \ref{tab:su3})
it is apparent that the potential for a molecular $Z_c(3900)$ and
$Z_{cs}(3985)$ are identical (provided their tentative identifications
with the $I^G(J^{PC}) = 1^+(1^{+-})$ $D^*\bar{D}$ and
$D^* \bar{D}_s$-$D \bar{D}_s^*$ systems are correct).
This in turn is compatible with their experimental masses,
as can be deduced from the qualitative argument
that their interaction is the same.
This conclusion has indeed been checked by concrete EFT
calculations~\cite{Yang:2020nrt},
which do not make hypotheses about the binding mechanism but simply assume
that the $Z_c$ and $Z_{cs}$ are bound states.
The question we will explore now is what are the conditions under which
we expect the OBE model to respect this SU(3)-flavor structure.

\begin{table}[!ttt]
\begin{tabular}{|cccc|}
\hline \hline
  System & $I$ & $S$ & $V$ \\
  \hline
  $D^{(*)}\bar{D}^{(*)}$ & 0 & 0 & $\frac{2}{3} V^{(S)} + \frac{1}{3} V^{(O)}$ \\
  $D_s^{(*)}\bar{D}_s^{(*)}$ & 0 & 0 &
  $\frac{1}{3} V^{(S)} + \frac{2}{3} V^{(O)}$ \\
  $D^{(*)}\bar{D}^{(*)}$ & 1 & 0 & $V^{(O)}$ \\
  $D^{(*)}\bar{D}_s^{(*)}$ & $\tfrac{1}{2}$ & -1 & $V^{(O)}$ \\
  \hline \hline 
\end{tabular}
\caption{SU(3)-flavor structure of the charmed meson-antimeson potential.
  The charmed mesons (antimesons) belong to the $\bar{3}$ ($3$) representation
  of SU(3)-flavor symmetry, from which the potential accepts a
  $3 \otimes \bar{3}  = 1 \oplus 8$ decomposition,
  which we show here.
  Notice that the SU(3)-flavor structure has to be combined with the HQSS
  structure in order to get the full S-wave potential,
  see Eqs.~(\ref{eq:V-singlet}) and (\ref{eq:V-octet}).
}
\label{tab:su3}
\end{table}

\subsection{Flavor structure of the light-meson exchanges}

When extending the present formalism from SU(2)-isospin to SU(3)-flavor,
a problem appears regarding the coupling of the charmed
and light mesons: the isoscalar ($I=0$) light-mesons,
being $q\bar{q}$ states, can be either in a flavor
singlet or octet configuration.
The singlet and isoscalar octet states, having the same quantum numbers,
can mix and this mixing most often works out as to separate
the $(u \bar{u} + d \bar{d}) / \sqrt{2}$ and $s\bar{s}$
components of these two type of
mesons almost perfectly.
This is what happens for instance with the vector mesons $\omega$ and $\phi$.
However, the other light mesons we are considering here are
further away from decoupling.
The easiest case will be the pseudoscalar mesons, for which the singlet
and octet almost do not mix.
The axial meson case will be the most complex one, as it entails non-trivial
mixing angles that have to be combined with the fact that the axial meson
mixes with the pion current.
In the following lines we will consider each case in detail.

\subsubsection{Pseudoscalar meson octet}

We will begin with the pseudoscalar mesons for which the singlet and octet
($\eta_1$ and $\eta_8$) members can be identified
with the $\eta'$ and $\eta$ mesons,
as the mixing angle is small.
Thus in practice we can consider the pseudoscalar mesons
as forming a standard octet  
\begin{eqnarray}
  M = \begin{pmatrix}
    \frac{\pi^0}{\sqrt{2}} + \frac{\eta}{\sqrt{6}} & \pi^{+} & K^+ \\
    \pi^- & -\frac{\pi^0}{\sqrt{2}} + \frac{\eta}{\sqrt{6}} & K^0 \\
    K^- & \bar{K}^0 & -\frac{2 \eta}{\sqrt{6}} 
  \end{pmatrix} \, . \label{eq:P8}
\end{eqnarray}
From this, the interaction term of the pseudoscalars
with the charmed mesons can be written as
\begin{eqnarray}
  \mathcal{L}_{\rm flavor} = \frac{g_1}{f_{\pi}}\,
          {\rm Tr}
          \left[ H_{\bar c}^{a\dagger} \vec{\sigma} \cdot \vec{\partial} M_{ab}
            H_{\bar c}^b
            \right]
          \, ,
\end{eqnarray}
where $a, b$ are flavor indices, which are ordered as
$\bar{D}^a = (\bar{D}^{0}, D^{-}, D_s^-)$, and $H_{\bar c}$
the heavy-superfield for the charmed antimesons
(as with this choice we have light-quarks).
Alternatively, if we use the light-subfield notation we will have
\begin{eqnarray}
  \mathcal{L}_{\rm flavor} = \frac{g_1}{f_{\pi}}\,
          q_L^{a\dagger} \vec{\sigma}_L \cdot \vec{\partial} M_{ab} q_L^b
          \, ,
\end{eqnarray}
with $q_L^a = (u_L, d_L, s_L)$ for $a = 1,2,3$

\subsubsection{Vector meson nonet}

Next we consider the vector mesons, for which the non-strange and strange
components of the singlet ($\omega_1$) and octet ($\omega_8$) decouple
almost perfectly to form the $\omega$ and $\phi$ mesons.
While the light-quark content of the singlet and isoscalar octet mesons
is expected to be
\begin{eqnarray}
  | \omega_1 \rangle &=&
  \frac{1}{\sqrt{3}}\,\left[ | u \bar{u} \rangle + | d \bar{d} \rangle
    + | s \bar{s} \rangle 
    \right] \label{eq:omega-1} \, , \\
  | \omega_8 \rangle &=& 
  \frac{1}{\sqrt{6}}\,\left[ | u \bar{u} \rangle + | d \bar{d} \rangle
    - 2 | s \bar{s} \rangle 
    \right] \label{eq:omega-8} \, ,
\end{eqnarray}
for the physical $\omega$ and the $\phi$ meson we have
\begin{eqnarray}
  | \omega \rangle &\simeq&
  \frac{1}{\sqrt{2}}\,\left[ | u \bar{u} \rangle + | d \bar{d} \rangle
    \right] \, , \\
  | \phi \rangle &\simeq& | s \bar{s} \rangle \, , 
\end{eqnarray}
which means that the relation between the physical and SU(3) eigenstates is
\begin{eqnarray}
  \begin{pmatrix}
    \phi \\
    \omega
  \end{pmatrix} &\simeq&
  \begin{pmatrix}
    \sqrt{\frac{1}{3}} & - \sqrt{\frac{2}{3}} \\
    \sqrt{\frac{2}{3}} & \sqrt{\frac{1}{3}}  
  \end{pmatrix}
  \,
  \begin{pmatrix}
    \omega_1 \\
    \omega_8
  \end{pmatrix} \, .
\end{eqnarray}
This matrix is actually a rotation, as can be seen by direct inspection.

In principle there should be two independent couplings for the singlet and
the octet vector meson components, i.e. $g_{V}^{(1)}$ for $\omega_1$ and
$g_{V}^{(8)}$ for $\omega_8$ and $\rho$ (or $f_{V}^{(1)}$ and $f_{V}^{(8)}$
for the magnetic-type couplings).
These two couplings are reduced to one once we consider the OZI rule:
the coupling of hadrons that do not contain strange quarks to
the $\phi$ meson should be suppressed.
This in turn generates a relation between $g_{V}^{(1)}$ and $g_{V}^{(8)}$
(there is only one independent coupling owing to the OZI rule),
from which we can deduce the relation $g_{\rho} = g_{\omega}$.

Alternatively, if we consider that the mixing is indeed ideal,
we can write down a vector meson nonet matrix
\begin{eqnarray}
  V =
  \begin{pmatrix}
    \frac{\rho^0}{\sqrt{2}} + \frac{\omega^0}{\sqrt{2}} & \rho^+ & K^{*+} \\
    \rho^- & -\frac{\rho^0}{\sqrt{2}} + \frac{\omega^0}{\sqrt{2}} & K^{*0} \\
    \bar{K}^{*-} & \bar{K}^{*0} & \phi
  \end{pmatrix} \, ,
\end{eqnarray}
and notice that the structure of the interaction Lagrangian
in flavor space will be
\begin{eqnarray}
  \mathcal{L}_{\rm flavor} \propto \bar{D}_a^{\dagger} V_{ab} \bar{D}_b \, ,
\end{eqnarray}
with $a, b$ flavor indices and $\bar{D}_a$ the anticharmed meson field
in flavor space, i.e. $\bar{D}_a = (\bar{D}^0 , D^0, D_s^{-} )$
for $a = 1, 2, 3$.
From this we end up with a unique $g_V$ and $f_V$, and
thus $g_{\rho} = g_{\omega}$ and $f_{\rho} = f_{\omega}$ automatically.

\subsubsection{Axial meson octet vs nonet}

The SU(3)-extension of the present formalism to the axial mesons
will encounter three problems.
The first is which are the flavor partners of the $a_1(1260)$ meson,
which we will simply assume to be the $f_1(1285)$, $f_1(1420)$ and
$K_1(1270)$ (we will further discuss this point later).

The second problem is the singlet and octet mixing:
if we consider that the isoscalar partners of the $a_1$ are the $f_1(1285)$
and $f_1(1420)$, they will be a non-trivial mixture of a singlet
and octet axial meson
\begin{eqnarray}
  \begin{pmatrix}
    f_1(1285) \\
    f_1(1420)
  \end{pmatrix} =
  \begin{pmatrix}
    \cos{\theta_{1}} & \sin{\theta_{1}} \\
   -\sin{\theta_{1}} & \cos{\theta_{1}}
  \end{pmatrix} \,
    \begin{pmatrix}
    f_1^1 \\
    f_1^8
  \end{pmatrix} \, , \label{eq:f1-mixing}
\end{eqnarray}
where $f_1^1$ and $f_1^8$ are the singlet and octet
components of the two $f_1$'s.
As a matter of fact, the $f_1(1285)$ and $f_1(1420)$ are relatively far away
from the mixing angle which effectively separates them into non-strange and
strange components.
If we define the decoupling mixing angle as
$\theta_{\rm dec} = {\rm atan}{( 1/\sqrt{2} )} \sim 35.3 {}^{\circ}$,
the $\theta_1$ angle can be expressed as
\begin{eqnarray}
  \theta_1 = \theta_{\rm dec} + \alpha_1 \, ,
\end{eqnarray}
where there is a recent determination of this angle by the LHCb,
$\alpha_1 = \pm (24.0^{+3.1}_{-2.3} \, {}^{+0.6}_{-0.8}) {}^{\circ}$~\cite{Aaij:2013rja} (which is the value we will adopt), and previously
in the lattice $\alpha_1 = \pm (31 \pm 2) {}^{\circ}$~\cite{Dudek:2011tt}.

Third, the axial neutral mesons are $J^{PC} = 1^{++}$ states but their strange
partners do not have well-defined C-parity.
Depending on their C-parity there are two possible types of axial mesons:
$J^{PC} = 1^{++}$ and $1^{+-}$ mesons originate from $^3P_1$ and $^1P_1$
quark-antiquark configurations, where we have used
the spectroscopic notation $^{S+1}L_J$, with $S$, $L$ and $J$ being
the spin, orbital and total angular momentum of the quark-antiquark pair.
For a quark-antiquark system the C-parity is $C = (-1)^{L+S}$,
which translates into $C=+1(-1)$ for $^3P_1$ ($^1P_1$).
The $J^{PC} = 1^{++}$ and $1^{+-}$ neutral axial mesons correspond with the
$a_1$, $f_1$ and $b_1$, $h_1$, respectively, of which only the $a_1$, $f_1$
can mix with the pseudo Nambu-Goldstone boson axial current.
The strange partners of the $a_1$ and $b_1$ axial mesons are referred to
as the $K_{1A}$ and $K_{1B}$, but for the strange axial mesons C-parity
is not a well-defined number and the physical states are a mixture of
the $^1P_1$ and $^3P_1$ configurations
\begin{eqnarray}
  \begin{pmatrix}
    K_1(1270) \\
    K_1(1400)
  \end{pmatrix} =
  \begin{pmatrix}
    \cos{\theta_K} & \sin{\theta_K} \\
   -\sin{\theta_K} & \cos{\theta_K}
  \end{pmatrix} \,
    \begin{pmatrix}
    K_{1B} \\
    K_{1A}
  \end{pmatrix} \, ,
\end{eqnarray}
with most determinations of $\theta_K$ usually close to either $30^{\circ}$
or $60^{\circ}$~\cite{Suzuki:1993yc,Burakovsky:1997dd,Roca:2003uk}.

The mixing of the axial pseudo Nambu-Goldstone meson current
(i.e. the SU(3) extension of the axial pion current)
has to happen with the axial meson octet
(instead of the physical axial mesons)
\begin{eqnarray}
  \partial_{\mu} M_{ab} \to \partial_{\mu} M_{ab} + \lambda_1 m_1 A^8_{1 ab} \, ,
\end{eqnarray}
where $M_{ab}$ and $A_{1 ab}$ would be the pseudoscalar and axial meson octets,
the first of which is given by Eq.~(\ref{eq:P8}) and the second by
\begin{eqnarray}
  A^8_1 = \begin{pmatrix}
    \frac{a_1^0}{\sqrt{2}} + \frac{f_1^8}{\sqrt{6}} & a_1^{+} & K_{1A}^{+*} \\
    a_1^- & -\frac{a_1^0}{\sqrt{2}} + \frac{f_1^8}{\sqrt{6}} & K_{1A}^{0*} \\
    K_{1A}^{-*} & \bar{K}_{1A}^{0*} & -\frac{2 f_1^8}{\sqrt{6}} 
  \end{pmatrix} \, .
\end{eqnarray}
Thus the specific relations for the contribution of
the $f_1^8$ and $K_{1A}$ to the axial meson currents, i.e.
\begin{eqnarray}
  \partial_{\mu} \eta &\to& \partial_{\mu} \eta + \lambda_1 m_{1} f_{1}^8 \, , \\
  \partial_{\mu} K &\to& \partial_{\mu} K + \lambda_1 m_{1} K_{1A} \, ,
\end{eqnarray}
have to be translated into the physical basis by undoing the rotations.
For the $f_{1}^8$ we will get
\begin{eqnarray}
  \partial_{\mu} \eta &\to& \partial_{\mu} \eta +
  \lambda_1 m_{1} \,
  \left( \sin{\theta_{1}} f_1 + \cos{\theta_{1}} f_1^* \right)
  \, , 
\end{eqnarray}
which determines the coupling of the $f_1$ and $f_1^*$ with
the $D$ and $D_s$ mesons, while for the $K_{1A}$ we will get
instead
\begin{eqnarray}
  \partial_{\mu} K &\to& \partial_{\mu} K +
  \lambda_1 m_{1} \left( \sin{\theta_K} K_1 + \cos{\theta_K} K_1^* \right)
  \, .
\end{eqnarray}
Owing to the form-factors, the exchange of the heavier variants of
the axial mesons ($f_1^*$, $K_1^*$) are expected to be suppressed
with respect to the lighter ones ($f_1$, $K_1$).
From this observation and the previous relations, the most important
contribution for axial-meson exchange will come from the couplings of
the $f_1$ and $K_1$ mesons, which are proportional to $\sin{\theta_1}$
and $\sin{\theta_K}$, respectively.

The $K_1$ meson deserves a bit more of discussion as it can help us to get
a sense of the accuracy of the previous relation from a comparison
with the $K_1$ axial meson decay constant, which can be extracted
from experimental information.
The current mixing relation implies that the decay constant will be
\begin{eqnarray}
  \langle 0 | A_{5 \mu} | K_1 \rangle = f_{K_1} m_{K_1} \quad \mbox{with} \quad
  f_{K_1} = \lambda_1 \cos{\theta_K} f_K \, , \nonumber \\
\end{eqnarray}
with from ${\theta_K} = 30-35^{\circ}$ or $55-60^{\circ}$
and $f_K = 160\,{\rm MeV}$ yields
$f_{K_1} = 110-150\,{\rm MeV}$ or $f_{K_1} = 180-230\,{\rm MeV}$,
respectively, which is to be compared
with $f_{K_1} = 175 \pm 19\,{\rm MeV}$~\cite{Cheng:2003bn} (which in turn
is extracted from the experimental data of Ref.~\cite{Barate:1999hj}).
The two possible mixing angles are in principle compatible with the previous
determination of $f_{K_1}$: though it is possible to argue that the higher
angle might be a slightly better choice, this is based on the assumption
that $f_{K1A} \neq 0$ but $f_{K1B} = 0$.
However, while the axial decay constant of the neutral $^1P_1$ axial mesons
$b_1$ and $h_1$ have to be zero owing to their negative C-parity,
i.e. $f_{b1} = 0$ and $f_{h1} = 0$, this is not true for the $K_{1B}$
for which $f_{K1B} = 0$ is a consequence of chiral symmetry.
In fact $f_{K1B} \neq 0$ owing to the finite strange quark mass.
This effect, though small, is enough as to make the comparison of
mixing angles more ambiguous~\cite{Cheng:2003bn,Cheng:2011pb}.

Finally, it is worth stressing that the structure of the axial mesons is
not particularly well-known and there exist interesting conjectures
about their nature in the literature.
A few hypotheses worth noticing are: (i) the axial mesons might be dynamically
generated (i.e. molecular)~\cite{Roca:2005nm,Geng:2008ag},
(ii) the $K_1(1270)$ resonance might actually have a
double pole structure~\cite{Geng:2006yb,Dias:2021upl}
and (iii) the $f_1(1420)$ might simply be a $K\bar{K}$
decay mode of the $f_1(1285)$~\cite{Debastiani:2016xgg}.
All of them might potentially influence the theoretical treatment of the axial
mesons: (i) actually was considered in Ref.~\cite{Durso:1977ek} four decades
ago for the $a_1(1260)$, where it was determined that it would not strongly
influence the form of the potential. It is worth noticing that (iii)
would imply that there are not enough axial mesons to form a nonet,
but only an octet.
This would be interesting, as in this scenario it might be plausible to identify
the $f_1(1285)$ with $f_1^8$ in Eq.~(\ref{eq:f1-mixing}),
leading to $\theta_1 = 90^{\circ}$.
However, though interesting, we will not consider the multiple ramifications of
the previous possibilities in this work.

\subsubsection{Scalar meson singlet vs nonet}

The lightest scalar meson nonet is formed by
the $\sigma$ (or $f_0(500)$), $a_0(980)$, $f_0(980)$ and $K_0^*(700)$.
If we are considering light-meson exchange the most important of the scalars
will be the lightest one, i.e. the $\sigma$ (see Ref.~\cite{Pelaez:2015qba}
for an extensive review about the status of this meson).
While the $a_0$ and $K_0^*$ are pure octets, the $\sigma$ and $f_0(980)$
are a mixture of singlet and octet, i.e.
\begin{eqnarray}
  \begin{pmatrix}
    f_0(500) \\
    f_0(980)
  \end{pmatrix} =
  \begin{pmatrix}
    \cos{\theta} & \sin{\theta} \\
   -\sin{\theta} & \cos{\theta}
  \end{pmatrix} \,
    \begin{pmatrix}
    S_1 \\
    S_8
  \end{pmatrix} \, ,
\end{eqnarray}
where $S_1$ and $S_8$ represent the pure singlet and octet states.
The meaning of $S_1$ and $S_8$ depends however on the internal structure of
the scalar mesons: if the $\sigma$ were to be a $q\bar{q}$ state, the
light-quark content of $S_1$ and $S_8$ would be analogous to that of
the vector mesons, i.e. to $|\omega_1 \rangle$ and $|\omega_8 \rangle$
in Eqs. (\ref{eq:omega-1}) and (\ref{eq:omega-8}).
But if the $\sigma$ were to be a $qq\bar{q}\bar{q}$ state, the light-quark
content of  $S_1$ and $S_8$ would be different (yet easily obtainable
from the substitutions $u \to \bar{d}\bar{s}$, $d \to \bar{u} \bar{s}$,
$s \to \bar{u}\bar{d}$, which assumes the diquark-antidiquark structure
proposed by Jaffe~\cite{Jaffe:1976ig}, where the antidiquark and diquark
are in a triplet and antitriplet configuration, respectively).

If $g_1$ and $g_8$ are the coupling of the charmed mesons to the singlet
and octet scalar, respectively, we will have that the coupling of
the $\sigma$ to the non-strange and strange charmed mesons will be
\begin{eqnarray}
  g_{\sigma 1} = g_{\sigma D D} &=& \cos{\theta}\,g_1 + \frac{1}{\sqrt{6}}\,\sin{\theta}\, g_8
  \, , \label{eq:mix-1} \\
  g_{\sigma 1}' = g_{\sigma D_s D_s} &=& \cos{\theta}\,g_1 - \frac{2}{\sqrt{6}}\,\sin{\theta}\, g_8
  \, . \label{eq:mix-2}
\end{eqnarray}
Independently of whether the $\sigma$ is a $q\bar{q}$ or $qq\bar{q}\bar{q}$
scalar, if we assume a mixing angle that decouples the non-strange and
strange components, we will end up with $g_{\sigma D_s D_s } = 0$ after
invoking the OZI rule (though we will discuss this point later).
In this case, $\sigma$ meson exchange will badly broke SU(3) symmetry
between the $Z_c(3900)$ and $Z_{cs}(3985)$.
However this conclusion depends on the previous assumptions,
which are not necessarily correct.
In the following lines we will discuss how the observed SU(3) symmetry
can still be preserved with scalar meson exchange.

The most obvious solution would be a flavor-singlet $\sigma$, as this
would provide roughly the same attraction for a molecular $Z_c(3900)$
and $Z_{cs}(3985)$.
In this regard it is relevant to notice Ref.~\cite{Oller:2003vf},
which analyzed the $\sigma$ pole in unitarized chiral perturbation theory
and obtained a mixing angle $\theta = 19 \pm 5{}^{\circ}$.
This would translate into a $\sigma$ that is mostly a flavor-singlet.

The interpretation of the $\sigma$ as a singlet would also be compatible
with the following naive extension of the L$\sigma$M from SU(2)-isospin
to SU(3)-flavor, in which an originally massless baryon octet
interacts with a total of nine bosons by means of
\begin{eqnarray}
  \mathcal{L}_{\rm int}^{\rm N \sigma L'} = g {\rm Tr} \, \left[
    \bar{B}_8 (\phi_0 + i\,\gamma_5 \lambda_a \phi_a ) B_8 \right] \, ,
  \label{eq:su3-lsigma}
\end{eqnarray}
with $B_8$ the baryon octet ($N$, $\Lambda$, $\Sigma$, $\Xi$),
$\phi_0$ and $\phi_a$ the bosonic fields, $\lambda_a$ with $a = 1, \dots, 8$
the Gell-Mann matrices and $g$ a coupling constant.
In the standard L$\sigma$M the nucleon field acquires mass owing to
the the spontaneous symmetry breaking and the subsequent vacuum
expectation value of the $\phi_0$ field.
Here it is completely analogous, with $\langle \phi \rangle = f_P / \sqrt{2}$,
the redefinition of $\phi_0 = f_P / \sqrt{2} + \sigma$ and
the reinterpretation of the $\phi_a$ fields as
the pseudoscalar octet ($\pi$, $K$, $\eta$).
This procedure will give $g_{\sigma B_8 B_8} = g_{\phi B_8 B_8} = \sqrt{2} M_8 / f_{P}$,
with $M_8$ the averaged mass of the octet baryons and $f_P$ representing
either $f_{\pi}$, $f_{K}$ or $f_{\eta}$, which are all identical
in the SU(3) symmetric limit. 
The $g_{\sigma B_8 B_8}$ thus obtained is basically compatible
with the previous SU(2) value for $g_{\sigma NN}$
($\simeq 10.2$).
Meanwhile the $F / (D+F)$-ratio would be $\alpha = 1/2$
(as the interaction term implies $F = D$), to be compared
for instance with the SU(6) quark model value $\alpha = 2/5$
(i.e. a $20\%$ discrepancy).

All this would suggest the use of approximately the same coupling
of the $\sigma$ to strange and non-strange hadrons alike,
resulting in the same attraction strength for both
the $Z_c(3900)$ and $Z_{cs}(3985)$.
In fact if we assume the relation $g_{\sigma q q} = \sqrt{2}\,m_q / f_P$ at the
quark level, take $m_q = 336$, $340$, $486\,{\rm MeV}$ for the $q = u, d, s$
constituent quark masses and choose $f_{P} = f_{\pi}$ for $q = u,d$ and
$f_{P} = f_K$ for $q=s$, we would get $g_{\sigma uu} \simeq g_{\sigma dd} \simeq 3.6$
and $g_{\sigma ss} \simeq 4.3$, leading to the counter-intuitive conclusion
that the coupling of the $\sigma$ to the strange quark is larger than
to the $u$, $d$ quarks.
However if we subtract the mass of the quarks $g_{\sigma q q} = \sqrt{2}\,(m_q^{\rm con} - m_{q}) / f_P$ (with $m_q^{\rm con}$ and $m_q$ the constituent and
standard quark masses), we will obtain $g_{\sigma ss} \simeq 3.4$ instead
(i.e. approximately identical to $g_{\sigma uu}$ and $g_{\sigma dd}$).

But the SU(3) extension of the L$\sigma$M we have presented here
is not the only possible one.
In fact it could just be considered as a simplified version of
the chiral quark model~\cite{Zhang:1997ny}
in which the scalar octet is removed.
It happens that the inclusion of the scalar octet in the chiral quark model
makes it perfectly possible to have a non-strange $\sigma$ and still
explain the mass of the light baryons.

However the problem might not necessarily be  whether the $\sigma$ contains
a sizable strange component or not, but whether it couples to
the strange degrees of freedom.
In this regard it has been suggested that the OZI rule does not
apply in the scalar $0^{++}$ sector~\cite{Isgur:2000ts,Meissner:2000bc}.
This in turn might be the most robust argument in favor of a sizable coupling
of the $\sigma$ meson with the strange-charmed mesons, as it does not depend
on the flavor structure or the strange content of the $\sigma$.
This in principle implies that the $g_{\sigma}$ and $g_{\sigma}'$ couplings
to the $D^{(*)}$ and $D_s^{(*)}$ charmed mesons would be independent
parameters.

Without the OZI rule there is no reason for the $g_1$ and $g_8$ couplings
to have comparable sizes: while the application of the OZI rule implies
that $g_8 = \sqrt{6}/2\,\tan{\theta}\,g_1$ (which for the $q\bar{q}$
and $qq\bar{q}\bar{q}$ ideal decoupling angles would translate
into $g_8 = \sqrt{3}/2 \, g_1 \simeq 0.87 \, g_1$ and
$g_8 = -\sqrt{3} \, g_1 \simeq -1.73 \, g_1$,
respectively), without OZI $g_1$ and $g_8$
are independent parameters.
Now, if it happens that $g_1 \gg g_8$ the result will be indistinguishable
from a flavor-singlet $\sigma$: the $g_{\sigma}$ and $g_{\sigma}'$ couplings
can be approximated by $g_{\sigma} \simeq g_{\sigma}' \simeq g_1 \cos{\theta}$,
resulting in approximately the same couplings to the $D^{(*)}$ and
$D_s^{(*)}$ charmed mesons.
As to whether the $g_1 \gg g_8$ condition is met or not,
it happens that $g_1 \cos{\theta}$ can be identified
with $g / 3$ in the SU(3)-extension of the L$\sigma$M of
Eq.~(\ref{eq:su3-lsigma}), giving it a relatively large value,
while there is no reason why $g_8$ should be as large.
Besides, $g_1 \gg g_8$ would also justify not including
the scalar octet in the OBE model.

Yet, we might get a better sense of the sizes of $g_1$ and $g_8$
from a comparison with previous determinations of
the $\sigma$ coupling in the light-baryon sector.
While the Nijmegen baryon-baryon OBE models originally considered
a flavor-singlet $\sigma$~\cite{Nagels:1976xq}, latter this idea
was put aside in favor of a more standard singlet-octet
interpretation for the $\sigma$~\cite{Nagels:1978sc}.
Their description depended on a singlet and octet couplings, $g_{1\,B_8 B_8}$
and $g_{8\,B_8 B_8}$, the mixing angle $\theta$ and the $F/(F+D)$-ratio
(which is necessary in the light-baryon octet
but not for the antitriplet charmed mesons).
They obtained $g_{8\, B_8 B_8} = 0.22\,g_{1\, B_8 B_8}$, while for the later
Nijmegen soft-core baryon-baryon OBE model~\cite{Maessen:1989sx}
this ratio is $g_{8\, B_8 B_8} = 0.34\,g_{1\, B_8 B_8}$.
Thus it would not be a surprise if the relation $g_1 \gg g_8$ also happens 
for the charmed mesons.

The comparison of the coupling constants to the light baryons
might provide further insights too.
If we consider the J\"ulich hyperon-nucleon OBE model~\cite{Holzenkamp:1989tq},
their results are $g_{\sigma \Lambda \Lambda} \simeq 0.95\,(0.77)\,g_{\sigma NN}$ and
$g_{\sigma \Sigma \Sigma} \simeq 1.13\,(1.05)\,g_{\sigma NN}$ in what is referred
to as model A(B) in Ref.~\cite{Holzenkamp:1989tq}, where these couplings
are supposed to represent correlated and uncorrelated (correlated)
processes in the scalar channel.
It is worth noticing that the J\"ulich model~\cite{Holzenkamp:1989tq} predated
the {\it rediscovery} of the $\sigma$ as a pole in the pion-pion
scattering amplitude~\cite{Dobado:1996ps,GarciaMartin:2011jx},
and consequently treated the $\sigma$ as a fictitious degree of freedom.
From a modern point of view in which the $\sigma$ is not a fictitious meson,
their results would be compatible (within the expected $30\%$ error
of the L$\sigma$M) both with a $\sigma$ that only couples
with the non-strange $q=u,d$ quarks
($g^{NS}_{\sigma \Lambda \Lambda} = 0.67\,g^{NS}_{\sigma NN}$ and
$g^{NS}_{\sigma \Sigma \Sigma} = 0.67\,g^{NS}_{\sigma NN}$, where $^{\rm NS}$ indicate that
it couples only to the non-strange quarks) and with a $\sigma$ that couples
with equal strength to the $q=u,d,s$ quarks
($g^{\rm FS}_{\sigma \Lambda \Lambda} = g^{\rm FS}_{\sigma NN}$ and
$g^{\rm FS}_{\sigma \Sigma \Sigma} = g^{\rm FS}_{\sigma NN}$,
where $^{\rm FS}$ indicates that the coupling is
flavor-symmetric).

In short, there are theoretical arguments in favor of a sizable coupling of
the $\sigma$ meson to the $D_s$ and $D_s^*$ charmed mesons, $g_{\sigma}'$.
In what follows we will consider the problem form a phenomenological
point of view, i.e. we will simply consider the $g_{\sigma}'$ coupling
to be a free parameter and discuss which are the values which allow
for a simultaneous description of the $Z_c$ and $Z_{cs}$,
without regard as to which is the theoretical reason
behind this.

\subsection{Light-meson exchange for the $Z_c$ and $Z_{cs}$}

Now that we have reviewed the flavor structure of the pseudoscalar, scalar,
vector and axial mesons, we can write down the resulting
light-meson exchange potential for the $Z_c$ and $Z_{cs}$.
We will begin with the pseudoscalar mesons, for which the singlet and octet
can be considered as effectively decoupled.
For the $Z_c$ there will be $\pi$- and $\eta$-exchange,
while for the $Z_{cs}$ only $\eta$-exchange
will be possible.
The pseudoscalar-exchange potential can be written as
\begin{eqnarray}
  V_{P}(D^*\bar{D}) &=&
  \zeta\,\vec{\tau}_1 \cdot \vec{\tau_2}\, W_{\pi}(r) +
  \frac{1}{3} W_{\eta}(r) \, , \\
  V_{P}(D_s^*\bar{D}) &=& -\frac{2}{3} \, W_{\eta}(r) \, , 
\end{eqnarray}
where $W_{\pi}$ and $W_{\eta}$ are the $\pi$- and $\eta$-exchange potentials
once we have removed the flavor and G-parity factors, i.e.
\begin{eqnarray}
  W_{\pi}(r) &=& \frac{g_1^2}{6 f_{\pi}^2} \,
  \vec{\sigma}_{L1} \cdot \vec{\sigma}_{L2} \,
  \mu_{\pi} W_Y(\mu_{\pi} r) \, , \\
  W_{\eta}(r) &=& \frac{g_1^2}{6 f_{\eta}^2} \,
  \vec{\sigma}_{L1} \cdot \vec{\sigma}_{L2} \,
  \mu_{\eta} W_Y(\mu_{\eta} r) \, .
\end{eqnarray}
In the flavor-symmetric limit we will have $m_{\pi} = m_{\eta}$ and
$f_{\pi} = f_{\eta}$, leading to identical $\pi$- and $\eta$-exchange
potentials, $W_{\pi} = W_{\eta}$.
However in the real world, $m_{\eta} > m_{\pi}$ and $f_{\eta} > f_{\pi}$,
implying a suppression of the $\eta$-exchange potential
relative to the pion-exchange one.
In particular, we will take $f_{\eta} = 150\,{\rm MeV}$.

For the vector mesons there is instead an almost ideal mixing
between the singlet and octet, from which the $\omega$ and $\phi$
are close to being purely non-strange and strange, respectively.
The structure of the potential will be
\begin{eqnarray}
  V_{V}(D^*\bar{D}) &=&
  \zeta\,\vec{\tau}_1 \cdot \vec{\tau_2}\, W_{\rho}(r) +
   W_{\omega}(r) \, , \\
  V_{V}(D_s^*\bar{D}) &=& 0 \, ,
\end{eqnarray}
with $W_{\rho}$ and $W_{\omega}$ the $\rho$- and $\omega$-exchange potential
once the flavor and G-parity factors have been removed
\begin{eqnarray}
  W_{\rho} &=& g_{\rho 1}^2 \, m_{\rho} W_Y(m_{\rho} r) \nonumber \\
    &+& f_{\rho 1}^2\frac{\mu_{\rho}^2}{6 M^2}\,\vec{\sigma}_{L1} \cdot \vec{\sigma}_{L2} \, \mu_{\rho} W_Y(\mu_{\rho} r) \, , \\
  W_{\rho} &=& g_{\omega 1}^2 \, m_{\omega} W_Y(m_{\omega} r) \nonumber \\
    &+&  f_{\omega 1}^2\frac{\mu_{\omega}^2}{6 M^2}\,\vec{\sigma}_{L1} \cdot \vec{\sigma}_{L2} \, \mu_{\omega} W_Y(\mu_{\omega} r) \, .
\end{eqnarray}

For the scalar meson we will consider it to generate two independent couplings
for the non-strange and strange charmed mesons:
\begin{eqnarray}
  V_{\sigma}(D^*\bar{D}) &=& - g_{\sigma 1}^2 \, m_{\sigma} W_Y(m_{\sigma} r) \, , \\
  V_{\sigma}(D_s^*\bar{D}) &=& - g_{\sigma 1}^{'} g_{\sigma 1} \, m_{\sigma} W_Y(m_{\sigma} r) \, , 
\end{eqnarray}
as this choice allows to explore the conditions for which we expect
the $Z_{cs}$ to bind, provided that the $Z_c$ binds.
We suspect that $g_{\sigma 1}$ and $g_{\sigma 1}'$ are of the same order of
magnitude, yet provided $|g_{\sigma 1}'|$ is not much smaller
than $|g_{\sigma 1}|$, the $Z_{c}$ and $Z_{cs}$
will be related to each other.

For the axial mesons, the $f_1(1285)$ and $f_1(1420)$ are probably
a non-standard mixture between a singlet and and octet, where
the mixing angles will have to be taken into account
explicitly
\begin{eqnarray}
  V_{A}(D^*\bar{D}) &=&
  \zeta\,\vec{\tau}_1 \cdot \vec{\tau_2}\,W_{a1}(r) \nonumber \\
  &+& \frac{1}{3} \Big[ \sin{\theta_1}^2 W_{f1}(r) \nonumber \\
   && \quad +
    \cos{\theta_1}^2 W_{f1^*}(r)    \Big]  \, , \\
  V_{A}(D_s^*\bar{D}) &=& -\frac{2}{3}
  \Big[ \sin{\theta_1}^2 W_{f1}(r) \nonumber \\ && \quad +
    \cos{\theta_1}^2 W_{f1^*}(r)    \Big]  \, .
\end{eqnarray}
The reduced $W_{a1}$, $W_{f1}$ and $W_{f1}^*$ potentials are given by
\begin{eqnarray}
 W_{a1} &=& -\lambda_1^2\,\frac{g_1^2 m_{a1}^2}{3 f_\pi^2}\,
 \vec{\sigma}_{L1} \cdot \vec{\sigma}_{L2} \, \mu_{a1} W_Y(\mu_{a1} r) \, , \\
 W_{f1} &=& -\lambda_1^2\,\frac{g_1^2 m_{f1}^2}{3 f_\eta^2}\,
 \vec{\sigma}_{L1} \cdot \vec{\sigma}_{L2} \, \mu_{f1} W_Y(\mu_{f1} r) \, ,
 \label{eq:W-f1} \\
 W_{f1}^* &=& -\lambda_1^2\,\frac{g_1^2 m_{f1}^{*2}}{3 f_\eta^2}\,
 \vec{\sigma}_{L1} \cdot \vec{\sigma}_{L2} \, \mu_{f1}^* W_Y(\mu_{f1}^* r) \, ,
 \label{eq:W-f1star}
\end{eqnarray}
with $m_{a1}$, $m_{f1}$ and $m_{f1}^*$ the masses of the $a_1$, $f_1$ and $f_1^*$
axial mesons (while $\mu_{a1}$, $\mu_{f1}$ and $\mu_{f1}^*$ are their effective
masses when there is a mass gap).
In general the exchange of the $f_1$ and $f_1^*$ mesons will be moderately
suppressed owing to $f_{\eta} > f_{\pi}$.

\subsection{The two $Z_{cs}$-like configurations}

There is an interesting difference between the isovector ($Z_c$) and
strange ($Z_{cs}$) sectors: in the first, $C$-parity is a good
quantum number for the neutral component of the isospin
triplet, i.e. for the $Z_c^0$, while in the second
this is not the case.
For the $D^* \bar{D}_s$-$D \bar{D}_s^*$ molecules, even if we consider
these two channels to be degenerate (which we do here), the structure of
the potential is still better understood
as a coupled channel problem, i.e.
\begin{eqnarray}
  V({Z_{cs}}) = \begin{pmatrix}
    V_a^{(1)} & V_b^{(1)} \\
    V_b^{(1)} & V_a^{(1)}
    \end{pmatrix} \, , \label{eq:V-Zcs-CC}
\end{eqnarray}
where $V_a^{(1)}$ and $V_b^{(1)}$ are the central and spin-dependent parts of
the potential, see Eqs.~(\ref{eq:V-light-spin}-\ref{eq:V1b}).
The two eigenvalues of the previous potential are:
\begin{eqnarray}
  V(\tilde{Z}_{cs}) &=& V_a^{(1)} + V_b^{(1)} \, , \label{eq:V-Zcs-tilde} \\
  V({Z}_{cs}) &=& V_a^{(1)} - V_b^{(1)} \, ,\label{eq:V-Zcs-standard}
\end{eqnarray}
which would be the strange counterparts of the $1^{++}$ and $1^{+-}$
isovector configurations in Table~\ref{tab:potential-XZ}.
In our C-parity convention the $\tilde{Z}_{cs}$ and $Z_{cs}$ wave functions
would be:
\begin{eqnarray}
  | \tilde{Z}_{cs} \rangle &=&
  \frac{1}{\sqrt{2}} \left[ | D^* \bar{D}_s \rangle +
    | D \bar{D}_s^* \rangle \right] \, , \label{eq:Zcs-tilde-wf} \\
  | {Z}_{cs} \rangle &=&
  \frac{1}{\sqrt{2}} \left[ | D^* \bar{D}_s \rangle -
    | D \bar{D}_s^* \rangle \right] \, ,  \label{eq:Zcs-wf}
\end{eqnarray}
respectively~\footnote{Meanwhile, in the alternative C-parity convention
  the sign of $V_b^{(1)}$ in Eq.~(\ref{eq:V-Zcs-CC}) changes and the same
  is true for the linear combinations in Eqs.~(\ref{eq:Zcs-tilde-wf})
  and (\ref{eq:Zcs-wf}). However the potentials in Eqs.~(\ref{eq:V-Zcs-tilde})
  and (\ref{eq:V-Zcs-standard}) will remain the same.
}.
The most standard interpretation of the $Z_{cs}(3985)$ observed by BESIII
is that it is indeed the strange partner of
the $Z_c(3900)$~\cite{Yang:2020nrt,Du:2020vwb}, that is,
what we have called the $Z_{cs}$ configuration
in Eqs.~(\ref{eq:V-Zcs-standard}) and (\ref{eq:Zcs-wf}).
This is the interpretation we will follow here.

However it is worth noticing that in the BESIII data~\cite{Ablikim:2020hsk}
the $Z_{cs}(3985) {}^-$ is seen in the $D^{*0} D_s^-$ channel,
while in the $D^0 D_s^{*-}$ what seems to be seen
is a broader structure at a lower mass.
This might be compatible with two nearby $Z_{cs}$ and $\tilde{Z}_{cs}$ poles
generating a constructive and destructive interference
in the $D^* \bar{D}_s$ and $D \bar{D}_s^*$ channels.
Alternatively, it might be a consequence of the production mechanism.
The point though is that the existence of two $Z_{cs}$ poles is not implausible
and that we are not completely sure of which one corresponds to the state
observed by BESIII.
Indeed we find that the existence of both the $Z_{cs}$ and $\tilde{Z}_{cs}$
poles is a likely outcome of our present approach~\footnote{We notice that
  it has also been suggested that the existence of two $Z_{cs}$
  states of similar mass might explain~\cite{Meng:2021rdg}
  the recently discovered $Z_{cs}(4000)^+$ by the LHCb~\cite{LHCb:2021uow}
  (in addition to the $Z_{cs}(3985)^-$ state observed
  by BESIII~\cite{Ablikim:2020hsk}).
  This possibility is intriguing, but we do not consider it here because of
  the large width of the $Z_{cs}(4000)^+$ ($\Gamma \sim 130\,{\rm MeV}$),
  which is an order of magnitude larger than the width of the $Z_{cs}(3985)^-$
  and thus difficult to explain if these two resonances
  were to be HQSS partners.
  This is not impossible though, as the $Z_{cs}$ and $\tilde{Z}_{cs}$ contain
  different linear combinations of the $D^* \bar{D}_s$ and $D \bar{D}_s^*$
  channels.
}.

\subsection{The $Z_c(3900)$ and $Z_{cs}(3985)$ cutoffs}

With the light-flavor structure of the OBE potential at hand,
we simply have to choose the parameters (couplings, masses and mixing angles),
compare the cutoffs for which the $Z_c(3900)$ and $Z_{cs}(3985)$ bind and
check whether they are compatible with each other.
This is analogous to what we have already done with the $X(3872)$ and
the $Z_c(3900)$, though now the focus is the preservation of
SU(3)-flavor symmetry, from which we expect
\begin{eqnarray}
  \frac{\Lambda(Z_{cs})}{\Lambda(Z_c)} \simeq 1.0 \, .
\end{eqnarray}
Of course this relation is approximate: HQSS and SU(3)-flavor violations
will generate deviations from this cutoff ratio.
We expect HQSS and SU(3)-flavor breaking effects to be of the order of
$15\%$ and $20\%$ (i.e. $\Lambda_{\rm QCD} / m_c$ and the difference between
$f_{\pi}$ and $f_K$), respectively, which add up to $25\%$
if we sum them in quadrature, i.e.
\begin{eqnarray}
  \frac{\Lambda(Z_{cs})}{\Lambda(Z_c)} = 1 \pm 0.25 = (0.75-1.25) \, .
\end{eqnarray}
For the SU(3)-flavor breaking, an extra factor to be taken into account is
the relative sizes of the $D$ and $D_s$ mesons, which are not necessarily
the same.
If we use the electromagnetic radii as a proxy of the matter radii,
although they have not been experimentally measured, there are
theoretical calculations: in Ref.~\cite{Hwang:2001th} they are estimated to be
$\sqrt{r^2_{e.m}} \sim 0.43$ and $0.35\,{\rm fm}$
for the $D$ and $D_s$, respectively.
This indicates that the strange charmed meson $D_s$ is $0.82$ the size of
its non-strange partner, from which it would be expectable for
the form-factor cutoff of a $D_s \bar{D}_s$ molecule to be
a $22\%$ larger than a $D \bar{D}$ molecule.
This figure is in fact compatible with the $f_K$ and $f_{\pi}$ ratio
we mentioned before, but indicates a bias in the flavor uncertainties:
the naive expectation will be a larger cutoff for the $Z_{cs}$
than for the $Z_c$.
The $D_s \bar{D}$ molecules would be in between, with deviations at the $10\%$
level expected for the cutoff (biased towards larger cutoffs),
from which we might revise the range of acceptable
cutoff ratios to
\begin{eqnarray}
  \frac{\Lambda(Z_{cs})}{\Lambda(Z_c)} \simeq 1.1 \pm 0.25 = (0.85-1.35) \, ,
  \label{eq:ratio-interval}
\end{eqnarray}
i.e. we have moved the expected central value from $1$ to $1.1$ to reflect
on the smaller size of the strange charmed mesons.
If we consider a $\sigma$ that does not couple with the strange charmed
meson $D_s^{(*)}$, for $n_P = 2$ the $Z_c$ and $Z_{cs}$ eventually bind
for large enough cutoffs, though the ratio is too large
\begin{eqnarray}
  \frac{\Lambda^{\sigma \rm (NS)}(Z_{cs})}{\Lambda^{\sigma \rm (NS)}(Z_c)}
  \Big|_{\theta_1^+}
  &=& 3.69^{+1.16}_{-1.06}\,(3.35-4.92) \, , \\
  \frac{\Lambda^{\sigma \rm (NS)}(Z_{cs})}{\Lambda^{\sigma \rm (NS)}(Z_c)}
  \Big|_{\theta_1^-}
  &=& 3.57^{+1.10}_{-1.00}\,(3.24-4.74) \, , 
\end{eqnarray}
where $\theta_1^{\pm} = \theta_{\rm dec} \pm \alpha_1$, from which it can
be appreciated that the dependence on the $\theta_1$ mixing angle is weak.
The central value, its error and the bands corresponds to
$m_{\sigma} = 550\,{\rm MeV}$, $g_{\sigma} = 3.4 \pm 1.0$ and
$m_{\sigma} = 450-600\,{\rm MeV}$, check Eq.~(\ref{eq:Lambda-X})
and the explanations following it.
If we instead consider a $\sigma$ that couples with the same strength to
the non-strange and strange quarks, i.e. a $\sigma$ with a
flavor-symmetric coupling, we will get instead the ratios
\begin{eqnarray}
  \frac{\Lambda^{\sigma \rm (FS)}(Z_{cs})}{\Lambda^{\sigma \rm (FS)}(Z_c)} \Big|_{\theta_1^+}
  &=& 1.04^{+0.10}_{-0.04}\,(1.00-1.06) \, , \\
  \frac{\Lambda^{\sigma \rm (FS)}(Z_{cs})}{\Lambda^{\sigma \rm (FS)}(Z_c)} \Big|_{\theta_1^-}
  &=& 1.06^{+0.14}_{-0.05}\,(1.00-1.09) \, , 
\end{eqnarray}
which are basically independent of $\theta_1$ and compatible
with Eq.~(\ref{eq:ratio-interval}).

\begin{figure}[ttt]
\begin{center}
  \includegraphics[width=9.2cm]{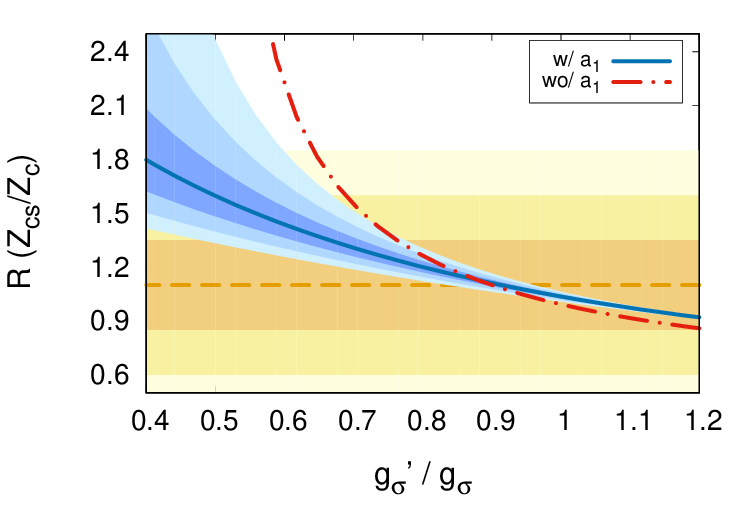}
\end{center}
\caption{
  Cutoff ratio between the $Z_{cs}(3985)$ and the $Z_c(3900)$ as a function
  of the ratio $g_{\sigma}' / g_{\sigma}$, where $g_{\sigma}$ and $g_{\sigma}'$
  are the couplings of the scalar meson to the non-strange and strange
  charmed baryons $D$ and $D_s$.
  We assume $g_{\sigma} = 3.4$, its expected central value
  from the L$\sigma$M and the quark model.
  We show the ratios in the OBE model with and without axial mesons,
  where we notice that for the axial-less case the $Z_{cs}'$
  does not bind for $g_{\sigma}' < 1.7$.
  We compare the cutoff ratio $R(Z_{cs}/Z_c)$ to the expected ratio derived
  from SU(3)-flavor, HQSS and corrections from the strange charmed meson
  size, $R(Z_{cs}/Z_c) \simeq 1.10 \pm 0.25$.
}
\label{fig:R-Zcs}
\end{figure}

The conclusion is that some coupling of the $\sigma$ meson to the strange
charmed meson is required for a coherent molecular
description of the $Z_c$ and $Z_{cs}$.
Thus we might consider the question of what is the $g_{\sigma}' / g_{\sigma}$
ratio which is compatible with the upper bound
for the cutoff ratio, i.e. with Eq.~(\ref{eq:ratio-interval}).
This happens to be
\begin{eqnarray}
  \frac{g_{\sigma}'}{g_{\sigma}} \geq  0.66-0.70 \, ,
\end{eqnarray}
which is weakly dependent of $m_{\sigma}$ (that is why we do not include
a bracket showing the $m_{\sigma}$ spread) and somewhat dependent on
$\theta_1$, with $\theta_1^{+}$($\theta_1^{-}$) yielding $0.66$ ($0.70$).
This ratio is compatible with a few of the different interpretations of
the $\sigma$ we have discussed: provided the $\sigma$ has a sizable
coupling to the strange components, it should be a plausible outcome.
For obtaining the cutoff ratio suggested by the strange and non-strange
charmed meson size comparison ($\simeq 1.1$) we will need instead
\begin{eqnarray}
  \frac{g_{\sigma}'}{g_{\sigma}} \simeq  0.91-0.94 \, ,
\end{eqnarray}
which again is nearly independent of $m_{\sigma}$ and weakly dependent on
$\theta_1$ ($\theta_1^{+}$($\theta_1^{-}$) gives $0.91$($0.94$)).
This ratio also falls within the realm of possibility,
but is more stringent.
The dependence of the cutoff ratio $R(Z_{cs} / Z_c)$ with
the $g_{\sigma}' / g_{\sigma}$ ratio is shown in Fig.~\ref{fig:R-Zcs},
from which can see again that axial meson exchange becomes important
if scalar meson exchange happens to be weak.
Owing to the weak dependence of this ratio on $\theta_1$,
Fig.~\ref{fig:R-Zcs} only shows the $\theta_1^{+}$ results

Finally, if we remove the sigma completely we will still get a ratio compatible
with Eq.~(\ref{eq:ratio-interval}),
\begin{eqnarray}
  \frac{\Lambda^{\slashed \sigma}(Z_{cs})}{\Lambda^{\slashed \sigma}(Z_c)} \Big|_{\theta_1^+}
  &=& 1.50 \, (1.33-4.19) \, , \\
  \frac{\Lambda^{\slashed \sigma}(Z_{cs})}{\Lambda^{\slashed \sigma}(Z_c)} \Big|_{\theta_1^-}
  &=& 1.45 \, (1.36-2.07) \, ,
\end{eqnarray}
where the intervals now reflect the uncertainty
{in $\lambda_1 = 1.8 \pm 0.3$}.
Sigma-less molecular descriptions include most works which use vector-meson
exchange (usually within the hidden-gauge approach) to predict molecular
states (e.g. the $X(3872)$~\cite{Gamermann:2007fi},
the hidden-charm pentaquarks~\cite{Wu:2010jy,Wu:2010vk}
or, recently, more general descriptions of
the molecular spectrum~\cite{Dong:2021juy}).
However these descriptions traditionally require a different binding mechanism
for the $Z_c(3900)$ resonance, which might include two-pion exchange or
charmonium exchange~\cite{Aceti:2014kja,Aceti:2014uea,He:2015mja,Dong:2021juy}.
Here we note that axial meson exchange could be a useful complementary
addition to these models, but if we want these models to simultaneously
reproduce the $X(3872)$ with the same set of parameters
the $\sigma$ is probably a required addition.

\subsection{The HQSS partners of the $Z_{cs}(3985)$}

Now we calculate the spectrum of the molecular partners of
the $Z_{cs}(3985)$ within the axial-full OBE model.
For simplicity we set the $\sigma$ coupling to the non-strange and
strange mesons to be identical, i.e. $g_{\sigma} = g_{\sigma}' = 3.4 \pm 1.0$,
and $\theta_1 = \theta_1^{+}$.
We use a dipolar form factor ($n_P = 2$) where the cutoff is obtained from
the condition of generating a pole in the $J^{P} = 1^{+}$
$(| D \bar{D}_s^* \rangle - | D^* \bar{D}_s \rangle) / \sqrt{2}$
scattering channel at threshold,
yielding $\Lambda({Z}_{cs}) = 1.88^{+1.02}_{-0.50}\,{\rm GeV}$.

The spectrum is shown in Table \ref{tab:molecules-strange}, where
it is worth noticing the following:
\begin{itemize}
\item[(i)] the details of the spectrum are less dependent on the strength
  of $\sigma$ exchange and there is already a considerable $D^* \bar{D}_s^*$
  hyperfine splitting for $g_{\sigma} = 3.4$.
  
\item[(ii)] However, the hyperfine splitting has the opposite sign to that
  of the $Z_c$ sector:
  \begin{itemize}
  \item[(ii.a)] In the $D^* \bar{D}_s$-$D \bar{D}_s^*$, the most attractive
    configuration is the $\tilde{Z}_{cs}$ one (instead of the $Z_{cs}$).
  \item[(ii.b)] The most attractive $D^* \bar{D}_s^*$ molecule
    is the $J^{P} = 2^{+}$ configuration (instead of the $0^{+}$ one).
  \end{itemize}
\end{itemize}
The reason for these features is pseudoscalar meson exchange:
$\eta$-exchange is considerably stronger in the $Z_{cs}$ sector
than in the $Z_c$ one, where at distances of the order of
$m_{\eta} r \sim 1$ we have
\begin{eqnarray}
  V_{P}(r; Z_c) &\approx& -\frac{g_1^2}{6}\,\left(
  \frac{m_{\pi}^2}{f_{\pi}^3 r} - \frac{1}{3} \frac{m_{\eta}^2}{f_{\eta}^3 r}
  \right)\,\vec{\sigma}_{L1} \cdot \vec{\sigma}_{L2} \, , \\
  V_{P}(r; Z_{cs}) &\approx& -\frac{g_1^2}{6}\,\left(
  \frac{2}{3} \frac{m_{\eta}^2}{f_{\eta}^3 r}
  \right)\,\vec{\sigma}_{L1} \cdot \vec{\sigma}_{L2} \, .
\end{eqnarray}
While in the flavor symmetric limit these two potentials would be identical,
when we input the physical pseudoscalar masses and decay constants
it becomes apparent that pseudoscalar meson exchange will be
much more important in the $Z_{cs}$ than in the $Z_c$
(actually, by a factor of $-2.8$, where the minus
sign is worth noticing).
In addition, $\eta$- and $\sigma$-exchange have approximately the same range,
which explains why it is not necessary to have a weak scalar coupling
in order to have a sizable hyperfine splitting.
Then, if we compare $\eta$-exchange with axial meson exchange,
each of which generate hyperfine splittings of the opposite sign,
the strength and range combination of the $\eta$ clearly
dominates over $f_1$ and $f_1^*$ meson exchanges.

This generates opposite hyperfine splitting patterns in the non-strange and
strange $D^*\bar{D}^*$ and $D^*\bar{D}_s^*$ sectors, a characteristic of
the spectrum that is unlikely to happen unless
the $Z_c$ and $Z_{cs}$ are molecular.
Depending on the nature of the $Z_c$ and $Z_{cs}$ resonances,
this prediction might be trivial to check:
for instance, if we assume all the $D^*\bar{D}^*$ and $D^*\bar{D}_s^*$
to be bound, the expected ordering of the masses of the states should be
\begin{eqnarray}
  && M_B(Z_c^*,0^{++}) < M_B(Z_c^*(4020)) < M_B(Z_c^*, 2^{++}) \, , \nonumber \\
  && M_B(Z_{cs}^*,0^{+}) > M_B(Z_{cs}^*, 1^{+}) > M_B(Z_{cs}^*, 2^{+}) \, ,
  \nonumber \\
\end{eqnarray}
where $M_B$ refers to the masses of the bound states.
On the other hand, if they all happen to be virtual states
the ordering would invert
\begin{eqnarray}
  && M_V(Z_c^*,0^{++}) > M_V(Z_c^*(4020)) > M_V(Z_c^*, 2^{++}) \, , \nonumber \\
  && M_V(Z_{cs}^*,0^{+}) < M_V(Z_{cs}^*, 1^{+}) < M_V(Z_{cs}^*, 2^{+}) \, ,
  \nonumber \\
\end{eqnarray}
with $M_V$ denoting their masses.
Unfortunately we do not know yet whether the $Z_c$ and $Z_{cs}$ are molecular,
less whether they are bound/virtual states or resonances above
the threshold~\cite{Albaladejo:2015lob,Yang:2020nrt}.
But independently of this, the eventual observation of the HQSS partners of
these two states, if accompanied by markedly different isospin splittings
in the $D^*\bar{D}^*$ and $D^* \bar{D}_s^*$ sectors, would indeed reveal
their molecular nature.

\begin{table}[!ttt]
\begin{tabular}{|c|cc|cc|}
\hline \hline
System ($Z_{cs}$-like) & $I$ & $J^{P}$ &
$B^{f_1}$ / $E_V^{f_1}$ & $M^{f_1}$ \\
\hline
$D\bar{D}_s$ & $\frac{1}{2}$ & $0^{+}$ &
$+1.5^{+0.1}_{-0.7}$ & $3834.1_{-0.1}^{+0.7}$\\
\hline
$\frac{1}{\sqrt{2}} (D^*\bar{D}_s + D \bar{D}_s^*)$ & $\frac{1}{2}$ & $1^{+}$ &
$+8.1^{+1.8}_{-6.5}$ & $3970.1^{+6.5}_{-1.8}$ \\
$\frac{1}{\sqrt{2}} (D^*\bar{D}_s - D \bar{D}_s^*)$ & $\frac{1}{2}$ & $1^{+}$ &
Input & Input \\
\hline
  $D^*\bar{D}_s^*$ & $\frac{1}{2}$ & $0^{+}$ &
  ${-1.8}^{+1.6}_{-1.2}$ & $4119.0^{+1.2}_{-1.6}$ \\
  $D^*\bar{D}_s^*$ & $\frac{1}{2}$ & $1^{+}$ &
  $+0.1^{+0.0}_{-0.0}$ & $4120.8^{+0.0}_{-0.0}$ \\
  $D^*\bar{D}_s^*$ & $\frac{1}{2}$ & $2^{+}$ &
  $+9.4^{+7.5}_{-2.4}$ & $4111.4^{+2.4}_{-7.5}$ \\
  \hline \hline
\end{tabular}
\caption{
  HQSS partners of the $Z_{cs}(3985)$ as a $D \bar{D}_s^*$-$D^*\bar{D}_s$
  molecule in the axial-full OBE model presented here.
  We use a dipolar form factor ($n_P = 2$) where the cutoff is determined
  from the condition of generating a pole at threshold in the
  $J^{P} = 1^{+}$ $D_s \bar{D}^*$-$D_s^*\bar{D}$ system,
  which for $g_{\sigma} = g_{\sigma}' = 3.4 \pm 1.0$ and $\theta_1 = \theta_1^{+}$
  results in $\Lambda({Z}_{cs}) = 1.88^{+1.02}_{-0.50}\,{\rm GeV}$.
  For the binding/virtual state energies we follow the same conventions
  as in Table \ref{tab:molecules}, where positive (negative) numbers
  indicate a bound (virtual) state.
  Masses and bound/virtual state energies are in units of ${\rm MeV}$.
}
\label{tab:molecules-strange}
\end{table}

\subsection{The scalar meson and the $P_{cs}(4459)$ pentaquark}

The strange and non-strange couplings of the scalar meson are not only
important for a unified molecular description of
the $Z_c(3900)$ and $Z_{cs}(3985)$,
but also for the new strange hidden-charm pentaquark
$P_{cs}(4459)$~\cite{Aaij:2020gdg} when considered in comparison to the other
three molecular pentaquark candidates: the $P_c(4312)$,
$P_c(4440)$ and $P_c(4457)$~\cite{Aaij:2019vzc}.

The $P_{cs}(4459)$ is $19.2\,{\rm MeV}$ below the $\bar{D}^* \Xi_c$ threshold
--- $4478.0\,{\rm MeV}$ in the isospin symmetric limit ---, which is why
the $P_{cs}(4459)$ has been conjectured to be a $\bar{D}^* \Xi_c$
molecule~\cite{Chen:2020uif,Peng:2020hql,Chen:2020kco,Liu:2020hcv,Zhu:2021lhd}.
The charmed baryon $\Xi_c$ is a flavor antitriplet 
with quark content $c s u$ ($\Xi_c^+$) and $c s d$ ($\Xi_c^0$), 
where the light-quark pair within the $\Xi_c$ is in a $S_L = 0$ configuration,
with $S_L$ the total light-quark spin.
As a consequence, pion exchange, axial meson exchange and the M1 part of
vector meson exchange do not contribute to the $\bar{D}^* \Xi_c$ interaction.
This observation also applies to the $\bar{D} \Sigma_c$ system~\cite{Chen:2019bip,Chen:2019asm,He:2019ify,Liu:2019tjn,Shimizu:2019ptd,Guo:2019kdc,Xiao:2019aya,Fernandez-Ramirez:2019koa,Wu:2019rog,Valderrama:2019chc}, which is
the most common molecular explanation for the $P_c(4312)$:
the $P_c(4312)$ pentaquark is merely $8.9\,{\rm MeV}$ below the
$\bar{D} \Sigma_c$ threshold.

The question is whether this is compatible with the expected binding of
the $P_{cs}(4459)$ as a $\bar{D}^* \Xi_c$ molecule.
Owing to the lack of explicit light-spin dependence, the only difference
between the OBE descriptions of the $P_c(4312)$ and $P_{cs}(4459)$ is scalar
meson exchange (the strength of vector meson being identical in both cases).
While the $\Sigma_c$ baryon contains two non-strange light-quarks,
the $\Xi_c$ contains only one and if we assume a $\sigma$ that
does not couple to the non-strange light-quarks we will have
\begin{eqnarray}
  g^{\rm NS}_{\sigma \Sigma_c \Sigma_c} \simeq \frac{1}{2}\,g^{\rm NS}_{\sigma \Xi_c \Xi_c}
  \, ,
\end{eqnarray}
which will translate into considerably less attraction (and binding)
for the $P_{cs}(4459)$.
In contrast, if the $\sigma$ couples with approximately
the same strength to the strange quark within the $\Xi_c$,
we will have
\begin{eqnarray}
  g^{\rm FS}_{\sigma \Sigma_c \Sigma_c} \simeq g^{\rm FS}_{\sigma \Xi_c \Xi_c}
  \, ,
\end{eqnarray}
where the superscript $^{\rm FS}$ indicates that the coupling is
now flavor-symmetric (in the sense of identical coupling
strengths with the $q = u, d, s$ quarks).
That is, if the $\sigma$ couples equally to all the light-quarks,
the binding of the $P_c(4312)$ and $P_{cs}(4459)$
molecules will be approximately the same.

However the experimental determination of the mass of the $P_{cs}(4459)$
indicates that it is more bound than the $P_c(4312)$ by about
$11.3\,{\rm MeV}$, where the specific binding energies are
\begin{eqnarray}
  B(P_c) = 8.9 \, {\rm MeV} \quad \mbox{and} \quad B(P_{cs}) = 19.2\, {\rm MeV}
  \, , \nonumber \\
\end{eqnarray}
for the $P_c(4312)$ and $P_{cs}(4459)$ pentaquarks, respectively.
This could be interpreted as $g_{\sigma \Xi_c \Xi_c} > g_{\sigma \Sigma_c \Sigma_c}$,
which would be somewhat surprising but still plausible.
Yet, the comparison we have done does not take into account that there is a
factor that generates spin-dependence in the $P_{cs}(4459)$ pentaquark:
the coupled channel dynamics with the nearby $\bar{D} \Xi_c'$ and
$\bar{D} \Xi_c^*$ thresholds, where the $J=1/2$ ($J=3/2$) $\bar{D}^* \Xi_c$
molecule will mix with the $\bar{D} \Xi_c'$ ($\bar{D} \Xi_c^*$) channel.
Owing to the relative location of the thresholds, this generates repulsion
and attraction for the $J=1/2$ and $3/2$ configurations, respectively.
Ref.~\cite{Peng:2020hql} computes this effect with a contact-range theory,
yielding
\begin{eqnarray}
  \Delta B^{CC} &=&
  B^{CC}(\bar{D}^{*} \Xi_c, J=\frac{1}{2}) -  B^{CC}(\bar{D}^{*} \Xi_c, J=\frac{3}{2}) \nonumber \\
  &\approx& (5-15)\,{\rm MeV} \, , 
\end{eqnarray}
depending on the assumptions made to calculate this effect, where
the superscript $^{CC}$ stands for ``coupled channels''.
This hyperfine splitting is of the same order of magnitude
as the aforementioned $11.3\,{\rm MeV}$ difference
in binding between the $P_c$ and $P_{cs}$.
For comparison,  a recent phenomenological calculation provides a similar
hyperfine splitting~\cite{Zhu:2021lhd} of
$\Delta B^{CC} = (2.4-20.0)\,{\rm MeV}$.

Yet, it is interesting to notice that
the LHCb collaboration~\cite{Aaij:2020gdg} already explored
the possibility that the $P_{cs}(4459)$ is actually composed
of two peaks instead of one.
For this two peak fit, the masses of the two $P_{cs}$ pentaquarks will be
$M(P_{cs1}) = 4454.9 \pm 2.7\,{\rm MeV}$ and
$M_2 = 4467.8 \pm 3.7\,{\rm MeV}$~\cite{Aaij:2020gdg}, yielding
the following binding energies
\begin{eqnarray}
  B(P_{cs1}) = 23.1 \, {\rm MeV} \quad \mbox{and} \quad
  B(P_{cs2}) = 10.2\, {\rm MeV}
  \, . \nonumber \\
\end{eqnarray}
We can describe this two-peak solution with the contact-range theory
of Ref.~\cite{Peng:2020hql}, which provides us with an
interesting advantage: we can explicitly turn off the coupled
channel effects within the theory to predict what the energy
of these two $P_{cs}$ pentaquarks would have been
in the absence of this effect, leading to
\begin{eqnarray}
  B^{SC}(\bar{D}^{*} \Xi_c, J=\frac{1}{2},\frac{3}{2})
  &=& (15.4-16.7)\,{\rm MeV} \, , 
\end{eqnarray}
with the superscript $^{SC}$ indicating ``single channel'' and
where now the two spin states are degenerate,
but have binding energies that are still somewhat larger
than the $P_c(4312)$ as a $\bar{D} \Sigma_c$ molecule.

The conclusion seems to be that, if the $P_{cs}(4459)$ pentaquark is a
$\bar{D}^* \Xi_c$ molecule (and assume that the previous procedures
effectively isolate the single and coupled channel contributions),
it is probably more compatible with a $\sigma$ that couples to all
the light quarks with equal strength than with a $\sigma$ that
does not couple with the strange quark.
If anything, it seems that there is more attraction in the $\bar{D}^* \Xi_c$
channel than in the $\bar{D} \Sigma_c$ one.
But this conclusion still depends on the size of the coupled channel effects
(they could have been underestimated) and the experimental uncertainties
surrounding the $P_{cs}(4459)$ pentaquark. 
Thus it might be possible that  the $P_{cs}(4459)$ pentaquark might
still be compatible instead with a sigma that does not couple to
the strange degrees of freedom.

\section{Conclusions}
\label{sec:conclusions}

In this manuscript we consider the problem of describing the $Z_{c}(3900)$,
$Z_c(4020)$ and $Z_{cs}(3985)$ as heavy hadron molecules
from a phenomenological perspective.
Of course, we do not know for sure whether they are molecular or not.
Instead, we are in interested in what their binding mechanism is
(provided they are molecular).
Regarding the problem of their nature, the closeness of these resonances to
the $D^*\bar{D}$, $D^*\bar{D}^*$ and $D^*\bar{D}_s$ thresholds suggest
a molecular nature.
The success of EFT formulations in describing the $Z_c$'s~\cite{HidalgoDuque:2012pq,Guo:2013sya,Albaladejo:2015lob} and $Z_{cs}$~\cite{Yang:2020nrt}
further points towards the plausibility of the molecular nature.
Yet, tetraquark explanations are also
possible~\cite{Chen:2010ze,Ali:2011ug,Lee:2008uy,Ferretti:2020ewe}.
What is not trivial to explain though in the molecular picture is
the binding mechanism: while the $Z_c$'s should not be there
in vector meson exchange models,
OBE models usually require relatively large cutoffs for these two-body
systems to bind~\cite{Liu:2008tn,Sun:2011uh,Liu:2019stu}
(or might simply not bind depending on the choice of couplings),
prompting other explanations such as two-pion exchange or charmonium
exchange~\cite{Aceti:2014kja,Aceti:2014uea,He:2015mja,Dong:2021juy}.

Here we consider a new factor in the molecular description of
the $Z_c$'s and $Z_{cs}$: axial meson exchange.
The exchange of axial mesons is strongly suppressed in the two-nucleon system,
partly owing to the fact that the axial meson mass is larger than the nucleon's
and partly owing to vector meson exchange being a more dominant factor
than the axial mesons.
But this is not necessarily the case for charmed mesons, prompting a
reevaluation of the role of axial mesons.
We find that the inclusion of the axial mesons makes the molecular description
more plausible for the $Z_c$'s, as they indeed provide additional attraction.
But their importance depends on the strength of scalar meson exchange:
if the coupling of the charmed hadrons to the scalar meson is smaller
than suggested by phenomenological models, axial mesons will become
the binding mechanism. Conversely, if the scalar coupling is large
enough, axial mesons will become irrelevant.
For molecular candidates in which vector meson exchange is strong, for instance
the $X(3872)$, the axial meson exchange contribution is negligible.
Thus we expect the relevance of axial meson exchange to be limited to
molecules where rho- and omega-exchange cancel out, as is
the case with the $Z_c$'s.

Yet, besides the axial meson, the nature of sigma exchange is probably
the most important factor for a coherent molecular description of
the $Z_c(3900)$ and $Z_{cs}(3985)$.
If the sigma meson couplings breaks SU(3)-flavor symmetry to a large degree,
the short-range of the axial meson, combined with its non-trivial
SU(3)-flavor structure, might be insufficient to explain the $Z_{cs}$
as the molecular SU(3)-partner of the $Z_c$
in the molecular picture.
Thus a molecular $Z_{cs}$ requires a non-trivial coupling of
the strange charmed mesons $D_s$ and $D_s^*$ with the sigma meson.
This is not improbable though, as there are theoretical reasons
(in particular the suspicion that the OZI rule might not apply
to the scalar mesons~\cite{Isgur:2000ts,Meissner:2000bc})
why the sigma meson could have a sizable coupling to
the strange degrees of freedom.
The bottom-line though is that a molecular $Z_{cs}$ requires a non-negligible
coupling of the sigma to the strange hadrons in the OBE model,
independently of which is the origin of this coupling.
This might not only be a requirement for the $Z_{cs}$ to be molecular
but also for the recently discovered $P_{cs}(4459)$,
the interpretation of which as a $\bar{D}^* \Xi_c$ bound state might
also require a coupling of the $\Xi_c$ strange charmed baryon
to the sigma similar to that of the non-strange $\Sigma_c$.

Finally, the observable signature of axial meson exchange will be a particular
type of hyperfine splitting for the isovector $D^*\bar{D}^*$ molecules,
where attraction decreases with spin.
If the isovector $D^*\bar{D}^*$ molecules are bound states or resonances,
there will be a $J^{PC} = 0^{++}$ $Z_c$-like state that is probably
$0-10\,{\rm MeV}$ lighter than the $Z_c(4020)$.
This pattern is reversed for their strange partners, owing to flavor symmetry
breaking effects that are specific to the molecular hypothesis.
Thus, besides the standard prediction of a $Z_{cs}(4120)$ $J^{P} = 1^{+}$
$D^* \bar{D}_s^*$ partner of the $Z_{cs}(3985)$~\cite{Yang:2020nrt},
here we expect the existence of a lighter (about $10\,{\rm MeV}$)
$J^P = 2^{+}$ partner of the $Z_{cs}^*$.
However if the $Z_c$'s and $Z_{cs}$ happen to be virtual
states~\cite{Albaladejo:2015lob,Yang:2020nrt}, the previous
patterns will invert or become difficult to recognize,
requiring a much more complex theoretical analysis to determine
which are the most attractive spin configurations.

\section*{Acknowledgments}

We would like to thank Feng-Kun Guo and Eulogio Oset for a careful reading
of the manuscript.
M.P.V. would also like to thank the IJCLab of Orsay, where part of this work
has been done, for its long-term hospitality.
This work is partly supported by the National Natural Science Foundation
of China under Grants No. 11735003 and 11975041, the Fundamental
Research Funds for the Central Universities and the Thousand
Talents Plan for Young Professionals.


%

\end{document}